\documentclass[sigconf]{acmart}

\renewcommand\footnotetextcopyrightpermission[1]{} 

\usepackage{url}
\usepackage{amsmath,amsfonts}
\usepackage{graphicx}
\usepackage{enumitem}
\usepackage{xcolor}
\usepackage{bm}

\newcommand{\nats}{\mathbb{N}}
\newcommand{\reals}{\mathbb{R}}

\newcommand{\nngreals}{\mathbb{R}_{\ge0}}

\newcommand{\nnints}{\mathbb{Z}_{\ge0}}

\renewcommand{\epsilon}{\varepsilon}

\newcommand{\calX}{\mathcal{X}}
\newcommand{\calY}{\mathcal{Y}}

\newcommand{\calN}{\mathcal{N}}

\newcommand{\tth}{^{(t)}}
\newcommand{\atth}{^{*(t)}}
\newcommand{\ttp}{^{(t,t')}}

\newcommand{\inv}{^{\!-1}}

\newcommand{\bmA}{\mathbf{A}}
\newcommand{\tbmA}{\tilde{\mathbf{A}}}

\newcommand{\bma}{\mathbf{a}}
\newcommand{\tbma}{\tilde{\mathbf{a}}}

\newcommand{\hO}{\hat{O}}
\newcommand{\hf}{\hat{f}}

\newcommand{\muA}{\mu_\bmA}

\newcommand{\Lam}{\Lambda}

\newcommand{\LamA}{\Lam_\bmA}

\newcommand{\PsiA}{\Psi_\bmA}

\newcommand{\textsyn}{\textsf}
\newcommand{\textdat}{\textsf}
\newcommand{\textutl}{\textsf}
\newcommand{\textano}{\textsf}

\definecolor{cmykblue}{cmyk}{1,1,0,0}
\newcommand{\colorB}[1]{\textcolor{black}{#1}}
\newcommand{\colorR}[1]{\textcolor{black}{#1}}

\newif\ifconferenceon\conferenceonfalse
\ifconferenceon
\newcommand{\conference}[1]{#1}
\newcommand{\arxiv}[1]{}
\else
\newcommand{\conference}[1]{}
\newcommand{\arxiv}[1]{#1}
\usepackage{balance}
\fi

\AtBeginDocument{%
  }

\begin{document}

\title{Designing a Location Trace Anonymization Contest}

\settopmatter{authorsperrow=4}
\author{Takao Murakami}
\authornote{The first author contributed to a design of the contest, 
datasets 
(Section~\ref{sub:contest_datasets} 
and Appendices~\ref{sec:synthesizer} to \ref{app:home_regions}), 
and experiments 
(Sections~\ref{sec:pre_exp}, \ref{sub:other_metrics}, and \ref{sub:utility_contest} 
and Appendix~\ref{app:fairness_contest}). 
The other authors contributed to a design of the contest and are ordered alphabetically.}
\affiliation{%
  \institution{AIST}
  \country{}
}

\author{Hiromi Arai}
\affiliation{%
  \institution{RIKEN}
  \country{}
}

\author{Koki Hamada}
\affiliation{%
  \institution{NTT}
  \country{}
}

\author{Takuma Hatano}
\affiliation{%
  \institution{NSSOL}
  \country{}
}

\author{Makoto Iguchi}
\affiliation{%
  \institution{Kii Corporation}
  \country{}
}

\author{Hiroaki Kikuchi}
\affiliation{%
  \institution{Meiji University}
  \country{}
}

\author{Atsushi Kuromasa}
\affiliation{%
  \institution{Data Society Alliance}
  \country{}
}

\author{Hiroshi Nakagawa}
\affiliation{%
  \institution{RIKEN}
  \country{}
}

\author{Yuichi Nakamura}
\affiliation{%
  \institution{SoftBank Corp.}
  \country{}
}

\author{Kenshiro Nishiyama}
\affiliation{%
  \institution{LegalForce}
  \country{}
}

\author{Ryo Nojima}
\affiliation{%
  \institution{NICT}
  \country{}
}

\author{Hidenobu Oguri}
\affiliation{%
  \institution{Fujitsu Limited}
  \country{}
}

\author{Chiemi Watanabe}
\affiliation{%
  \institution{Tsukuba University of Technology}
  \country{}
}

\author{Akira Yamada}
\affiliation{%
  \institution{KDDI Research, Inc.}
  \country{}
}

\author{Takayasu Yamaguchi}
\affiliation{%
  \institution{Akita Prefectural University}
  \country{}
}

\author{Yuji Yamaoka}
\affiliation{%
  \institution{Fujitsu Limited}
  \country{}
}

\renewcommand{\shortauthors}{Murakami et al.}

\begin{abstract}
For a better understanding of anonymization methods for location traces, 
we have designed and held a \textit{location trace anonymization contest} 
that 
deals with a long trace 
(400 events per user) 
and fine-grained locations 
(1024 regions). 
In our contest, each team anonymizes her original traces, and then the other teams perform privacy attacks against the anonymized traces. 
\colorB{In other words, both defense and attack compete together, which is close to what happens in real life. 
Prior to our contest, 
we show} 
that re-identification alone is insufficient as a privacy risk and that trace inference should be added as an additional risk. 
Specifically, we show an example of anonymization that is \textit{perfectly} secure against re-identification and is not secure against trace inference. 
Based on this, our contest evaluates both the re-identification risk and trace inference risk and analyzes their relationship. 
Through our contest, we show several findings in a situation where both defense and attack compete together. 
In particular, we show that an anonymization method secure against trace inference is also secure against re-identification under the presence of appropriate pseudonymization. 
\colorB{We also report 
defense and attack algorithms that won first place, and analyze the utility of anonymized traces submitted by teams in various applications such as POI recommendation and geo-data analysis.} 
\end{abstract}

\keywords{location privacy, 
contest, re-identification, trace inference}

\maketitle
\pagestyle{plain}

\section{Introduction}
\label{sec:intro}
Location-based services (LBS) such as POI (Point of Interest) search, 
route finding, \colorB{and POI recommendation \cite{Cheng_IJCAI13,Feng_IJCAI15,Liu_VLDB17}} 
have been 
widely used in recent years. 
\colorB{For example, a smartphone or in-car navigation system may repeatedly send a user's location to an LBS provider to receive services such as route finding and POI recommendation. 
The sequence of locations is called a location trace.} 
With the spread of such services, 
a large amount of location traces 
are accumulating in a data center of the LBS provider. 
These traces 
can be provided to a third party 
to perform various geo-data analysis tasks 
such as mining popular POIs \cite{Zheng_WWW09}, auto-tagging POI categories (e.g., restaurants, hotels)~\cite{Do_TMC13,Ye_KDD11}, and modeling human mobility patterns 
\cite{Liu_CIKM13,Song_TMC06}.

However, the 
disclosure 
of location traces raises a serious privacy concern \colorB{(on 
leaks of sensitive data)}. 
For example, there is a risk that the traces are used to 
infer 
sensitive social relationships 
\cite{Bilogrevic_WPES13,Eagle_PNAS09} 
or hospitals visited by users. 
Some studies have shown that even if the traces are pseudonymized, original user IDs can be re-identified with high probability  \cite{Gambs_JCSS14,Mulder_WPES08,Murakami_PoPETs17}. 
This fact indicates that the pseudonymization alone is not sufficient and  
anonymization is necessary 
before providing 
traces 
to a third party. 
In the context of location traces, anonymization consists of location obfuscation (e.g., adding noise, generalization, and deleting some locations) and pseudonymization. 

Finding an appropriate anonymization method for location traces is extremely challenging. 
For example, each location in a trace can be highly correlated with the other locations. 
Consequently, an anonymization mechanism based on differential privacy (DP) \cite{Dwork_ICALP06,DP} 
\colorR{might not} 
guarantee high privacy and utility \colorB{(usefulness of the anonymized data for LBS or geo-data analysis)}. 
More specifically, 
\colorR{consider releasing location traces \cite{Andres_CCS13,Meehan_AISTATS21,Xiao_CCS15} or aggregate time-series (time-dependent population distribution) \cite{Pyrgelis_PoPETs17}.} 
It is well known that DP with a small privacy budget $\epsilon$ (e.g., $\epsilon = 0.1$ or $1$ \cite{DP_Li}) provides strong privacy against adversaries with arbitrary background knowledge. 
However, 
\colorR{DP might 
result in 
no meaningful privacy guarantees 
or poor utility 
for long traces  \cite{Andres_CCS13,Pyrgelis_PoPETs17} 
(unless we limit the adversary's background knowledge \cite{Meehan_AISTATS21,Xiao_CCS15}), as $\epsilon$ in DP tends to increase with increase in the trace length}. 
Thus, 
there is still no conclusive 
answer on how to appropriately anonymize location traces, which motivates our objective of designing a contest.

\smallskip
\noindent{\textbf{Location Trace Anonymization Contest.}}~~To better understand 
anonymization methods for 
traces, we have designed and held a location trace anonymization contest 
called PWS (Privacy Workshop) Cup 2019 \cite{PWSCup2019}. 
Our contest has two phases: \textit{anonymization phase} and \textit{privacy attack phase}. 
In the anonymization phase, 
each team anonymizes her original traces. 
Then in the privacy attack phase, the other teams obtain the anonymized traces and perform privacy attacks against the anonymized traces. 
Based on this, we evaluate 
privacy and utility of the anonymized traces for each team. 
In other words, we model the problem of finding an appropriate anonymization method as a 
\colorB{battle} 
between a 
designer of the anonymization method 
(each team) 
and adversaries (the other teams). 
One promising feature of our contest is that \textit{both defense and attack compete together}, which is close to what happens in real life. 

The significance of a location trace anonymization contest is not limited to finding appropriate anonymization methods. 
For example, it is useful for an \textit{educational purpose} 
-- 
everyone can join the contest to understand the importance of anonymization (e.g., why the pseudonymization alone is not sufficient) or to improve her own anonymization skill. 

However, designing a location trace anonymization contest poses great challenges. 
\colorB{In particular, an in-depth understanding of privacy risks is crucial for finding better anonymization methods. 
Below, we explain this issue in more detail.}

\smallskip
\noindent{\textbf{Privacy Risks.}}~~\colorB{Privacy risks have been extensively studied over the past two decades \cite{PPDM,Torra_book}. 
In particular, re-identification 
(or de-anonymization) 
has been acknowledged as a major risk in the privacy literature. 
Re-identification refers to matching anonymized data with publicly available information or auxiliary data to discover the individual to which the data belongs \cite{EC_re-identification}. 
In the literature on location privacy, re-identification is defined as a problem of mapping anonymized or pseudonymized traces to the corresponding user IDs \cite{Gambs_JCSS14,Mulder_WPES08,Murakami_PoPETs17,Shokri_SP11}. 
It has been demonstrated that re-identification attacks are very effective for apparently anonymized data, such as medical records in the Group Insurance Commission (GIC) \cite{Sweeney_UFKS02} and the Netflix Prize dataset \cite{Narayanan_SP08}. 
Re-identification of data subjects is also acknowledged as a major risk in data protection laws, such as General Data Protection Regulation (GDPR) \cite{GDPR,GDPR_iapp}.} 

\colorB{Attribute inference, 
which 
infers attributes of an individual, is another major privacy risk. 
In particular, 
it is known that some attributes may be 
inferred \textit{without} re-identifying records \cite{Torra_book,NISTIR8053}. 
Its famous example is 
an attribute inference attack against $k$-anonymity 
\cite{Machanavajjhala_ICDE06}. 
$k$-anonymity guarantees that every record is indistinguishable from at least $k-1$ other records with respect to quasi-identifiers 
(e.g., ZIP code and birth date). 
Thus, an adversary who knows only quasi-identifiers cannot re-identify a record with a probability larger than $1/k$. 
However, if the $k$ records have the same sensitive attribute (e.g., cancer), an adversary who knows a patient's quasi-identifiers can infer her sensitive attribute without re-identifying her record. 
This attack is called the homogeneity attack 
and motivates the notion of $l$-diversity 
\cite{Machanavajjhala_ICDE06}. 
Similarly, Shokri \textit{et al.} \cite{Shokri_SP11} show that $k$-anonymity for location traces is vulnerable to attribute inference attacks, where the attribute is a location in this case.}

\colorB{In this paper, we provide stronger evidence that re-identification alone is not sufficient as a privacy risk -- 
we provide 
an example of anonymization that is \textit{perfectly} secure against re-identification and is not secure against attribute inference. 
Specifically, we show that the \textit{cheating anonymization}~\cite{Kikuchi_AINA16} (or \textit{excessive anonymization} \cite{Nojima_JIP18}), which excessively anonymizes each record, has this property.}

\begin{figure}[t]
\centering
\includegraphics[width=0.99\linewidth]{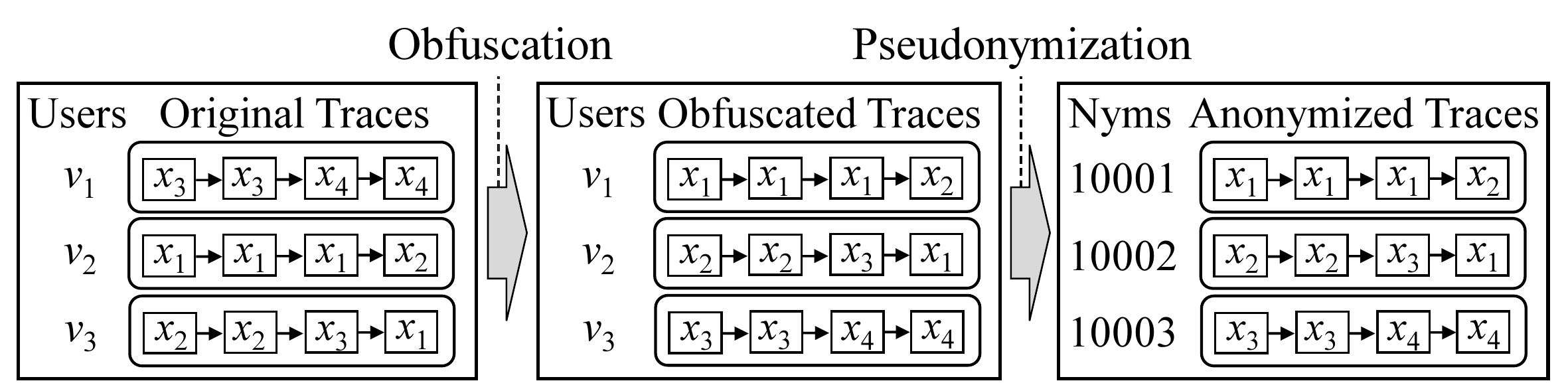}
\conference{\vspace{-2mm}}\arxiv{\vspace{-2mm}}
\caption{Cheating anonymization.}
\label{fig:cheat}
\end{figure}

In 
location traces, 
the cheating anonymization 
can be explained as follows. 
Consider a dataset 
in Figure~\ref{fig:cheat}. 
In this example, there are three users ($v_1$, $v_2$, and $v_3$) 
and four types of discrete locations ($x_1$, $x_2$, $x_3$, and $x_4$). 
We excessively add noise to their original traces so that obfuscated traces of users 
$v_1$, $v_2$, and $v_3$ are the same as the original traces of users $v_2$, $v_3$, and $v_1$, respectively. 
Note that this is \textit{location obfuscation} rather than pseudonymization. 
Then we pseudonymize these traces so that pseudonyms 
$10001$, $10002$, and $10003$ 
correspond to $v_1$, $v_2$, and $v_3$, respectively. 

This anonymization is seemingly 
insecure 
because it just shuffles the original traces in the same way as pseudonymization. 
However, this anonymization is secure against re-identification\footnote{\colorR{This is 
the reason that this anonymization is called ``cheating'' anonymization.}}, 
unlike pseudonymization. 
\colorR{To show this, 
consider an adversary who 
knows 
the original traces 
or any other traces highly correlated with the original traces.} 
This adversary would re-identify 
$10001$, $10002$, and $10003$ 
as $v_2$, $v_3$, and $v_1$, respectively. 
However, 
$10001$, $10002$, and $10003$ 
actually correspond to $v_1$, $v_2$, and $v_3$, respectively, as explained above. 
Therefore, this re-identification attack fails -- 
the re-identification rate is 
$\frac{0}{3}$. 
This is caused by the fact that we excessively obfuscate the locations so that the shuffling occurs before pseudonymization. 
Note that this shuffling can be viewed as a \textit{random permutation of user IDs before pseudonymization}. 
Thus, the cheating anonymization is \textit{perfectly} secure against re-identification in that the anonymized traces provide no information about user IDs; i.e., the adversary cannot find which permutation is correct. 

\colorB{Although the security of the cheating anonymization against re-identification is explained in \cite{Kikuchi_AINA16,Nojima_JIP18}, the security of this method against attribute inference has not been explored. 
Therefore, prior to our contest, we evaluate the privacy of the cheating anonymization against attribute inference through experiments.} 
Our experimental results 
show that 
the cheating anonymization 
is not secure against a \textit{trace inference attack} (a.k.a. 
\textit{tracking attack} \cite{Shokri_SP11}), which infers the whole locations in the original traces. 
In other words, 
we show 
that the adversary can 
recover the original traces from anonymized traces 
without 
re-identifying them. 
\colorB{Note that $k$-anonymity is not perfectly secure against re-identification unless $k$ is equal to the total number of records. 
In contrast, the cheating anonymization is perfectly secure against re-identification, as explained above. 
\colorR{Thus, our experimental results 
strongly 
demonstrate 
that 
re-identification alone is insufficient as a privacy risk and that trace inference should be added 
as an additional risk}\footnote{We also note that our experimental results are totally different from the vulnerability of mix-zones \cite{Bindschaedler_NDSS12}. 
The mix-zone is a kind of pseudonymization (not location obfuscation) that assigns many pseudonyms to a single user. 
Thus, it is vulnerable to re-identification, as explained in \cite{Bindschaedler_NDSS12}.}.}

\smallskip
\noindent{\textbf{\colorB{Our Contest.}}}~~\colorB{Based on 
our experimental results, 
our contest evaluates 
both the re-identification risk and trace inference risk and analyzes their relationship. 
In particular, since we know that the security against re-identification does not imply the security against trace inference, 
we pose the following question: 
\textit{Does the security against trace inference imply the security against re-identification?} 
Through our contest, we show that 
the answer is yes 
under the presence of appropriate pseudonymization.} 

We also show 
other findings based on our contest. 
First, we explain 
how 
\colorB{the best defense and attack algorithms that won first place in our contest} 
are effective compared to 
existing algorithms such as \cite{Bettini_Springer09,Chow_SIGKDD11,Gedik_TMC08,Gambs_JCSS14,Mulder_WPES08,Murakami_PoPETs17,Murakami_TIFS17,Shokri_SP11}. Second, we show that anonymized traces submitted by teams are useful for various 
applications such as POI recommendation \cite{Cheng_IJCAI13,Feng_IJCAI15,Liu_VLDB17} and geo-data analysis. 

\smallskip
\noindent{\textbf{Our Contributions.}}~~Our contributions are as follows: 

\begin{itemize}
    \item \colorB{Through experiments, we show that 
    \textbf{there exists an algorithm perfectly secure against re-identification and not secure against trace inference}.} 
    \item \colorB{Based on our experimental results, 
    we design and hold a location trace anonymization contest that evaluates both the re-identification and trace inference risks. 
    Through our contest, we 
    show that 
    \textbf{an anonymization method secure against trace inference is also secure against re-identification under the presence of appropriate pseudonymization}. 
    This finding is important because it provides a guideline on how to anonymize traces so that they are secure against both re-identification and attribute (trace) inference. 
    We also report 
    the best defense and attack algorithms in our contest 
    and 
    analyze their utility in various 
    applications such as POI recommendation and geo-data analysis.} 
\end{itemize}

\smallskip
\noindent{\textbf{Basic Notations.}}~~Let $\nats$, $\nnints$, and $\reals$ be the set of natural numbers, non-negative integers, and real numbers, respectively. 
For $a\in\nats$, let $[a] = \{1, 2, \cdots, a\}$. 
Let $\tau \in \nats$ be the number of teams in a contest. 
For $t \in [\tau]$, let $P_t$ be the $t$-th team. 
We use these notations throughout this paper.

\section{Related Work}
\label{sec:related}

\noindent{\textbf{\colorB{Re-identification and Attribute Inference.}}}~~\colorB{Re-identification has been 
acknowledged as a major risk in the privacy literature and data protection laws, such as GDPR \cite{GDPR,GDPR_iapp}. 
Famous examples of re-identification attacks include Sweeney's attack against medical records \cite{Sweeney_UFKS02} and Narayanan-Shmatikov's attack against the Netflix Prize dataset \cite{Narayanan_SP08}.
Re-identification has also been studied in location privacy \cite{Gambs_JCSS14,Mulder_WPES08,Murakami_PoPETs17,Montjoye_SR13}. 
For example, de Montjoye \textit{et al.} \cite{Montjoye_SR13} show that three (resp.~four) locations in a trace are enough to uniquely characterize about $80\%$ (resp. $95\%$) of users amongst one and a half million users. 
This indicates that we need to sacrifice the utility (e.g., delete almost all locations from a trace) to prevent re-identification in the maximum-knowledge attacker model \cite{Domingo-Ferrer_PST15}, where the adversary knows the entire original traces as background knowledge. 
Some studies \cite{Gambs_JCSS14,Mulder_WPES08,Murakami_PoPETs17} show that re-identification 
still poses a threat 
in the partial-knowledge attacker model  \cite{Ruiz_PSD18,Murakami_TDP21}, where the adversary does not know the entire original traces. 
} 

\colorB{The relationship between re-identification and attribute inference has also been studied in the literature. 
As explained in Section~\ref{sec:intro}, the homogeneity attack \cite{Machanavajjhala_ICDE06} against $k$-anonymity is one of the most famous examples of ``attribute inference without re-identification.'' 
The vulnerability of $k$-anonymity to location inference is also shown in \cite{Shokri_SP11}. 
We strengthen these results by showing the existence of algorithms \textit{perfectly} secure against re-identification and not secure against trace inference.} 

\colorB{There is also a trivial example of ``re-identification without attribute inference'' \cite{Torra_book}. 
Specifically, let us consider 
the maximum-knowledge attacker 
who knows all attributes of users. 
This attacker can easily re-identify the users, as explained above. 
However, the attacker does not infer or newly obtain any attribute, as she already knows all attributes. 
In other words, 
we cannot evaluate the attribute inference risk 
in the maximum-knowledge attacker model.}

\colorB{In addition, although the maximum-knowledge attacker model is the worst-case model, it is unrealistic and overly pessimistic. 
Thus, we follow a threat model in \cite{Gambs_JCSS14,Mulder_WPES08,Pyrgelis_PoPETs17,Pyrgelis_NDSS18,Shokri_SP11} and separate the background knowledge of the adversaries from the original traces, i.e., partial-knowledge attacker model. 
In this model, 
it is unclear whether the security against attribute (trace) inference implies the security against re-identification. 
Thus, 
we explore this question through a contest where both defense and attack compete together.} 

\smallskip
\noindent{\textbf{Anonymization Contest.}}~~We also 
note that some anonymization contests have been held over a decade \cite{Jordon_arXiv20,Kikuchi_AINA16,Kikuchi_DPMQASA16,NIST_DPChallenge18,NIST_DPChallenge20}. 
The contests in \cite{Jordon_arXiv20,Kikuchi_AINA16,Kikuchi_DPMQASA16,NIST_DPChallenge18} do not deal with location traces \colorR{(\cite{Kikuchi_AINA16,Kikuchi_DPMQASA16,NIST_DPChallenge18} use microdata, and \cite{Jordon_arXiv20} uses clinical data)}. 
From October 2020 to June 2021 (after our contest in 2019), NIST held the Differential Privacy Temporal Map Challenge \cite{NIST_DPChallenge20}. 
In this contest, participants compete for a 
DP algorithm 
for 
a sequence of location events over three sprints 
using public datasets. 
This contest deals with a small sequence of events per individual (e.g., $7$ events in sprint 2) or coarse-grained 
locations 
(e.g., $78$ 
regions 
in sprint 3).

Our contest substantially differs from 
this contest 
in that 
participants anonymize 
a long trace ($400$ events per individual) with fine-grained locations ($1024$ regions) in our contest. 
In this case, \colorR{it is 
very 
difficult to 
release long traces under DP with a small $\epsilon$}
\cite{Andres_CCS13,Pyrgelis_PoPETs17}, as described in Section~\ref{sec:intro}. 
Therefore, we 
measure the privacy via the accuracy of re-identification or trace inference as in \cite{Gambs_JCSS14,Murakami_PoPETs17,Shokri_SP11}. 
\colorB{Our contest is also different from \cite{NIST_DPChallenge20} in that ours evaluates both the re-identification and trace inference risks and 
analyzes the relationship between the two risks.} 

\section{Design of a Location Trace Anonymization Contest}
\label{sec:contest}
This section explains 
the design of our location trace anonymization contest. 
First, Section~\ref{sub:purpose_approach} explains the purpose of our contest. 
Then, Section~\ref{sub:overview_contest} describes the overview of our contest. 
\colorB{Section~\ref{sub:contest_datasets} explains datasets used in our contest.} 
Finally, Section~\ref{sub:details_contest} describes the details of our contest. 

\subsection{Purpose}
\label{sub:purpose_approach}
As described in Section~\ref{sec:intro}, a location trace anonymization contest is useful for \textit{technical} 
and \textit{educational} purposes. 
\colorB{The main technical purpose is to 
find better 
trace anonymization methods 
in terms of privacy and utility. 
For privacy risks, 
two types of information disclosure are known: \textit{identity disclosure} and \textit{attribute disclosure} \cite{Torra_book}. 
Identity disclosure is caused by re-identification \cite{Torra_book}, which finds a mapping between anonymized/pseudonymized traces and user IDs in the case of location traces. 
Attribute (location) disclosure is caused by trace inference \cite{Shokri_SP11}, which infers the whole locations in the original traces from anonymized traces. 
Therefore, it is desirable for trace anonymization methods to have security against both re-identification and trace inference.} 

Taking this into account, 
we pose the following questions 
for a technical purpose: 
\begin{description}[leftmargin=8.75mm]
    \item[\colorB{RQ1.}]  Is an anonymization method that has security against re-identification also secure against trace inference?
    \item[\colorB{RQ2.}]  Is an anonymization method that has security against trace inference also secure against re-identification?
\end{description}

In Section~\ref{sec:pre_exp}, we show that the answer to the first question RQ1 is \textit{not} always yes by showing a counterexample -- the cheating anonymization \cite{Kikuchi_AINA16} is perfectly secure against re-identification but not secure against trace inference. 

Thus, the remaining question is the second one RQ2. 
Note that 
if we do not appropriately pseudonymize traces, we can find 
a trivial counterexample 
for this. 
As an extreme 
example, 
assume that we 
delete all (or almost all) 
locations in the original traces and pseudonymize  each trace so that a pseudonym includes the corresponding user ID (e.g., pseudonyms of $v_1$ and $v_2$ are ``10001-$v_1$'' and ``10002-$v_2$'', respectively). 
Clearly, such anonymization is 
secure against trace inference but not secure against re-identification. 

However, 
finding an answer to 
RQ2 
becomes \textit{non-trivial} 
when we appropriately pseudonymize 
(randomly shuffle) 
traces. 
\colorB{If the answer to RQ2 is yes, 
it has a significant implication for trace anonymization. 
Specifically, it provides a guideline on \textit{how to anonymize traces so that they are secure against the two attacks} -- 
one promising approach 
is 
to make traces secure against trace inference 
because it also implies security against re-identification.} 
However, 
empirical evidence (or a counterexample) for 
the yes-answer to RQ2 has not been 
established in the literature, 
especially in a situation where both defense and attack compete together.

Thus, we design our contest to find an answer to 
RQ2 under the presence of appropriate pseudonymization. 

\subsection{Overview of Our Contest}
\label{sub:overview_contest}
\begin{figure}[t]
\centering
\includegraphics[width=0.99\linewidth]{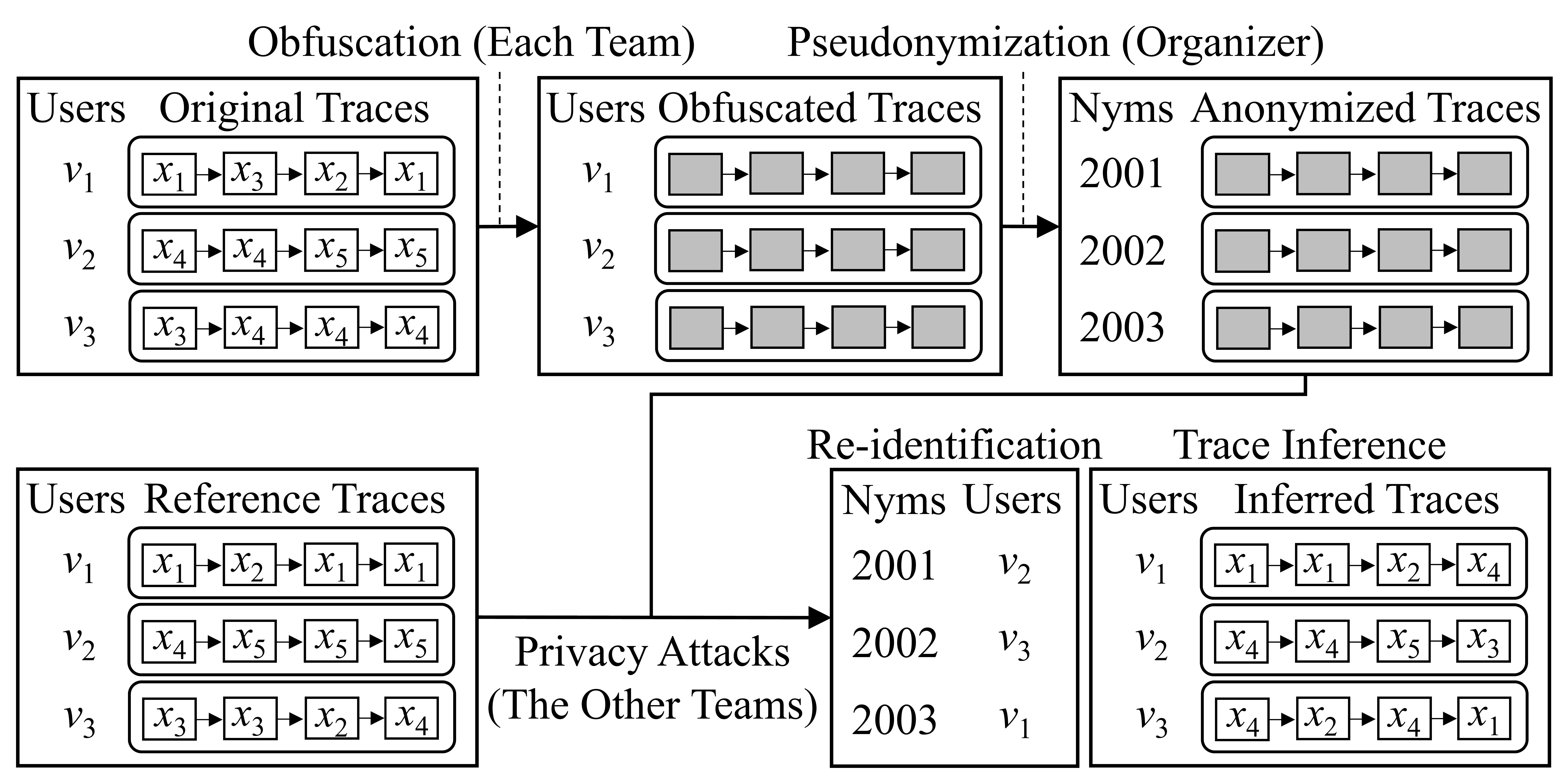}
\conference{\vspace{-2mm}}\arxiv{\vspace{-2mm}}
\caption{Overview of our contest. Gray squares represent that locations are obfuscated.}
\label{fig:contest_overview}
\end{figure}

We design our contest to achieve the purpose explained above. 
Figure~\ref{fig:contest_overview} shows its overview. 

First of all, an organizer in our contest performs pseudonymization (which is important but technically trivial) in place of each team to guarantee that pseudonymization is appropriately done. 
Thus, each team obfuscates traces and sends the traces to the organizer. 
Then the organizer pseudonymizes 
the traces, i.e., randomly shuffles the traces and adds pseudonyms. 

Each team obfuscates its original traces 
so that its anonymized traces are secure against \textit{trace inference} while keeping 
high utility. 
Then the other teams attempt \textit{both re-identification and trace inference} against the anonymized traces. 
\colorB{Here, we follow a threat model in the previous work \cite{Gambs_JCSS14,Mulder_WPES08,Pyrgelis_PoPETs17,Pyrgelis_NDSS18,Shokri_SP11}, where the adversary's background knowledge is separated from the original traces. 
Specifically,} 
the other teams 
perform 
the re-identification and trace inference 
attacks using \textit{reference traces}, 
which are separated from the original traces. 
\colorB{Note that they do not know the original traces, i.e., the partial-knowledge attacker model \cite{Ruiz_PSD18,Murakami_TDP21}.} 
The reference traces are, for example, traces of the same users on different days. 
\colorB{The users may disclose the reference traces via geo-social network services, or the adversary may obtain the reference traces by observing the users in person.} 
The reference traces play a role as background knowledge of the adversary. 

For each team, we evaluate the following three scores: \textit{utility score}, 
\textit{re-identification privacy score}, 
and 
\textit{trace inference privacy score}. 
Every score takes a value between $0$ and $1$ (higher is better). 
We regard an anonymized trace as \textit{valid} (resp.~\textit{invalid}) if its utility score is larger than or equal to (resp.~below) a pre-determined threshold. 
For an invalid trace, we set its privacy scores 
to $0$. 

Then we give 
the \textit{best anonymization award} to a team that achieves the highest 
trace inference privacy score\footnote{We 
also gave an award to a team that achieved the highest 
re-identification privacy score. 
Specifically, we distributed two sets of location data for each team: one for a \textit{re-identification challenge} and another for a \textit{trace inference challenge}. 
In the re-identification (resp.~trace inference) challenge, each team competed together to achieve the highest 
re-identification (resp.~trace inference) privacy score, 
and the winner got an award. 
We omit 
the re-identification challenge in this paper.}. 
We also give the \textit{best re-identification (resp.~trace inference) award} to a team that contributes the most in lowering 
re-identification (resp.~trace inference) privacy scores 
for the other teams. 

The best anonymization award motivates each team to anonymize traces so that they are secure against trace inference. 
The other two awards motivate each team to make every effort to attack them in terms of both re-identification and trace inference. 
Consequently, we can see whether anonymized traces intended to have security against trace inference are also secure against re-identification. 

\smallskip
\noindent{\textbf{Remark 1.}}~~Note that 
each team competes for privacy while satisfying the utility requirement in our contest. 
Each team does not compete for utility while satisfying the privacy requirement, because we would like to investigate the relationship between two privacy risks: re-identification and trace inference. 

\smallskip
\noindent{\textbf{Remark 2.}}~~It is also possible to design 
a contest that releases aggregate location time-series (time-dependent population distributions) \cite{Pyrgelis_NDSS18}. 
Our contest 
follows a general location privacy framework in \cite{Shokri_SP11} and 
releases anonymized traces. 
This is because the anonymized traces 
are useful 
for 
a very wide range of applications 
such as POI recommendation \cite{Liu_VLDB17} and geo-data analysis, as shown in 
Section~\ref{sub:utility_contest}. 

\smallskip
\noindent{\textbf{\colorR{Remark 3.}}}~~\colorR{As described in Section~\ref{sec:related}, the attacker can be divided into two types: \textit{partial-knowledge attacker} \cite{Ruiz_PSD18,Murakami_TDP21} and 
\textit{maximum-knowledge attacker} \cite{Domingo-Ferrer_PST15}. 
Clearly, the maximum-knowledge attacker 
is stronger than the partial-knowledge attacker. 
Our contest focuses on the partial-knowledge attacker model 
for two reasons. 
First, the maximum-knowledge attacker model is overly pessimistic; e.g., we need to sacrifice the utility to prevent re-identification in this model \cite{Montjoye_SR13}. 
Second, 
we cannot evaluate the attribute inference risk in the maximum-knowledge attacker model, as described in Section~\ref{sec:related}.} 

\colorR{Regarding privacy metrics, we can consider the following two types, depending on how to quantify the total risk: \textit{average metrics} and \textit{worst-case metrics}. 
The average metrics consider an average risk over all users, and their examples include the re-identification rate \cite{Gambs_JCSS14,Mulder_WPES08,Murakami_PoPETs21} and the adversary's expected error \cite{Shokri_SP11}. 
In contrast, the worst-case metrics consider a worst-case risk, and their examples include DP \cite{DP} and its variants \cite{Desfontaines_PoPETs20}. 
The worst-case metrics are stronger than the average metrics.}

\colorR{As will be explained later, our contest uses 
the average metrics -- our re-identification and trace inference privacy scores are based on the re-identification rate and the adversary's expected error, respectively. 
This is because the worst-case metrics such as DP might result in no meaningful privacy guarantees or poor utility 
for long traces \cite{Andres_CCS13,Pyrgelis_PoPETs17}, as described in Section~\ref{sec:intro}. 
Our contest considers DP out of scope in that it does not use DP as a privacy metric. 
We also note that we do not 
systematically evaluate actual privacy-utility trade-offs of existing DP algorithms when releasing long traces, 
which could be an interesting avenue for future work.}

\subsection{\colorB{Datasets}}
\label{sub:contest_datasets}

\noindent{\textbf{\colorB{Dataset Issue in the Partial-Knowledge Attacker Model.}}}~~\colorB{As described in Section~\ref{sub:overview_contest}, our contest assumes the partial-knowledge attacker model, where the adversary does not know the original traces and has reference traces separated from the original ones. 
In this case, it is challenging to prepare a contest dataset.} 

\colorB{Specifically, the challenge in the partial-knowledge attacker model is that public datasets (e.g., \cite{Geolife,Gowalla,SNS_people_flow,Yang_WWW19}) 
cannot be directly used for a contest. 
This issue comes from the fact that 
everyone can access public datasets. 
In other words, if we use a public dataset for a contest, then every team would know the original traces of the other teams, which leads to the maximum-knowledge attacker model. 
It is also difficult to directly use a private dataset in a company due to 
privacy concerns.} 

\colorB{One might think that the organizer can use a public dataset by setting a rule that submissions (obfuscated traces) must not include knowledge from the public dataset. 
However, this is an unrealistic solution because we cannot detect the rule violation. 
Specifically, we cannot verify whether or not submissions include knowledge from the public dataset, unless we accurately\footnote{\colorB{To ensure that the rule violation does not occur, the accuracy needs to be almost 100\%.}} extract information about the public dataset from the submissions, e.g., membership inference \cite{Jayaraman_USENIX19,Shokri_SP17}. 
The same applies to the case when the submissions are machine learning models, hyperparameters, or architectures -- they can be trained from a public dataset to provide better classification accuracy \cite{Bergstra_JMLR12, Elsken_JMLR19}, and we cannot detect the rule violation unless we accurately extract the public dataset from them.} 

\colorB{One might also think that the organizer can use a public dataset without announcing which public dataset it is. 
This 
is also unrealistic for two reasons. 
First, the number of public datasets is limited. 
Therefore, each team can easily obtain the public dataset corresponding to reference traces of other teams. 
In other words, it is very difficult (or impossible) to hide which public dataset is used. 
Second, we cannot detect the rule violation (i.e., the use of the public dataset corresponding to reference traces), as explained above.}

\colorB{A lot of existing work \cite{Bindschaedler_SP16,Chen_CCS12,Chen_KDD12,Chow_WPES09,He_VLDB15,Kido_ICPS05,Murakami_PoPETs21,You_MDM07} proposes location synthesizers, which take location traces (called \textit{training traces}) as input and output synthetic traces. 
However, they cannot be used for our contest. 
Specifically, most of them \cite{Chen_CCS12,Chen_KDD12,Chow_WPES09,He_VLDB15,Kido_ICPS05,You_MDM07} generate synthetic traces based on parameters common to all users and do not provide \textit{user-specific features}; e.g., someone lives in Manhattan, and another one commutes by train. 
Note that the user-specific features are necessary for an anonymization contest because otherwise, the adversary cannot re-identify traces. 
In other words, the adversary needs some user-specific features as background knowledge to re-identify traces.} 

\colorB{A handful of existing synthesizers \cite{Bindschaedler_SP16,Murakami_PoPETs21} preserve the user-specific features. 
However, both \cite{Bindschaedler_SP16} and \cite{Murakami_PoPETs21} generate traces so that 
the $i$-th synthetic trace preserves the user-specific feature of the $i$-th training trace. 
Thus, if the organizer uses a public dataset as training traces and discloses synthetic traces, each team can obtain the corresponding training traces in the public dataset. 
Note that even if the organizer shuffles synthetic traces or generates multiple traces per training trace, each team may link each synthetic trace with the corresponding training trace via re-identification to obtain better background knowledge. 
For example, it is reported in \cite{Murakami_PoPETs21} that the synthetic traces in \cite{Bindschaedler_SP16} can be easily re-identified (re-identification rate $=80$ to $90\%$) when they preserve statistical information about the training traces. 
In addition, the synthesizers in \cite{Bindschaedler_SP16,Murakami_PoPETs21} do not provide strong theoretical 
privacy guarantees such as DP \cite{Dwork_ICALP06,DP} when a private dataset 
is used as training traces.} 

\colorB{Therefore, neither public datasets, private datasets, nor existing location synthesizers can be used for our contest.}

\begin{figure}[t]
\centering
\includegraphics[width=0.95\linewidth]{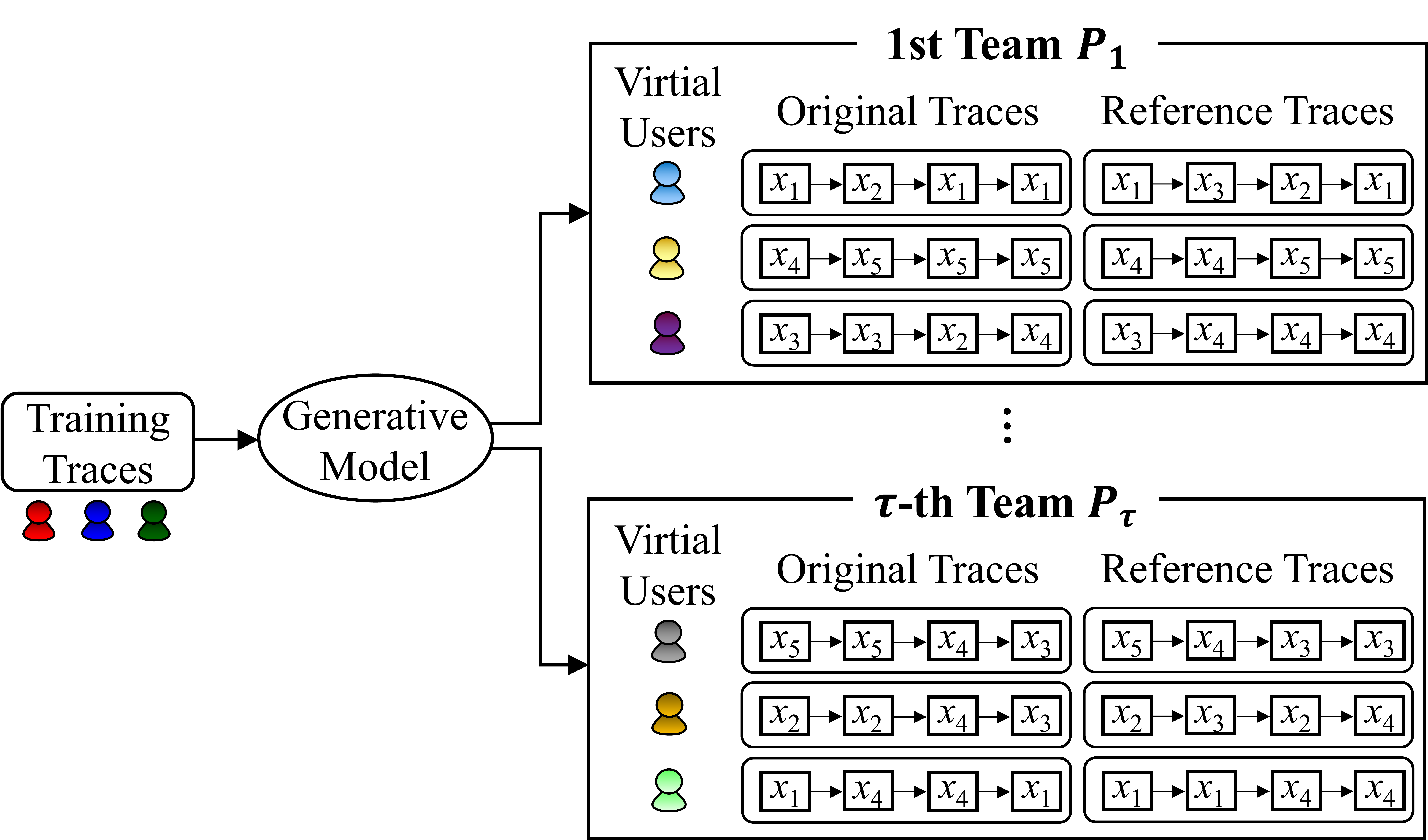}
\conference{\vspace{-3mm}}\arxiv{\vspace{-3mm}}
\caption{\colorB{Overview of our location synthesizer.}}
\label{fig:synthesizer_brief}
\end{figure}

\smallskip
\noindent{\textbf{\colorB{Location Synthesizer with Diversity.}}}~~\colorB{To address the dataset issue explained above, we introduce a location synthesizer that takes 
training traces 
as input and outputs different synthetic traces for each team. 
Figure~\ref{fig:synthesizer_brief} shows its overview. 
In a nutshell, our location synthesizer 
randomly generates traces of 
\textit{virtual users} who are different from users in the input dataset (called \textit{training users}), as indicated by different colors in Figure~\ref{fig:synthesizer_brief}. 
The virtual users for each team are also different from the virtual users for the other teams.} 

\colorB{In our location synthesizer, each virtual user has her own user-specific feature (e.g., 
live in Manhattan, commute by train) represented as a multi-dimensional vector. 
We call it a \textit{feature vector}. 
We synthesize each user's trace based on the feature vector. 
Consequently, each team's synthetic traces have different features than those of training traces and the other teams' synthetic traces. 
We call this \textit{diversity} of synthetic traces. 
The diversity prevents 
the $t$-th team $P_t$ from linking the other teams' reference traces with training traces or 
$P_t$'s 
original traces to obtain better background knowledge. 
The organizer does not disclose each team's original traces or feature vectors to the other teams. 
Thus, each team does not know the original traces of the other teams, i.e., the partial-knowledge attacker model.} 

\colorB{We extend the location synthesizer in \cite{Murakami_PoPETs21}, which preserves various statistical features of training traces (e.g., distribution of visit-fractions \cite{Do_TMC13,Ye_KDD11}, time-dependent population distribution \cite{Zheng_WWW09}, and transition matrix \cite{Liu_CIKM13,Song_TMC06}), to have diversity explained above. 
See Appendix~\ref{sec:synthesizer} for details of our location synthesizer. 
In Appendix~\ref{sec:exp_syn}, we show through experiments that our synthesizer has diversity and preserves various statistical features. 
In Appendix~\ref{app:privacy_location_synthsizer}, we also empirically evaluate the privacy of our location synthesizer 
when a private dataset is used as training traces\footnote{\colorB{The caveat of our location synthesizer, as well as the existing synthesizers in \cite{Bindschaedler_SP16,Murakami_PoPETs21}, is that it does not provide strong theoretical privacy guarantees such as DP. 
However,} 
in our contest, 
we used a public dataset \cite{SNS_people_flow} as training traces to synthesize traces. 
Thus, the privacy issue did not occur.}.} 

\colorB{In our contest, we slightly modified our synthesizer in Appendix~\ref{sec:synthesizer} to make synthetic traces more realistic in that users tend to be at their home between 8:00 and 9:00 and 17:00 and 18:00 while keeping statistical information (e.g., population distribution, transition matrix). 
See Appendix~\ref{app:home_regions} for details. 
We also published our synthesizer as open-source software \cite{LocSyn}.} 

\begin{figure}[t]
\centering
\includegraphics[width=0.99\linewidth]{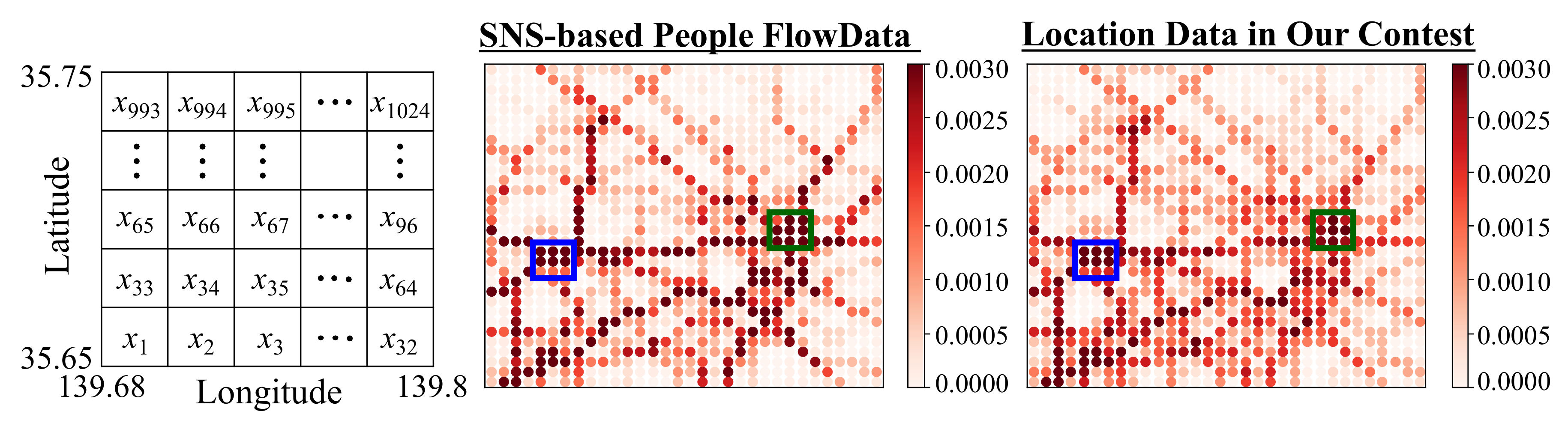}
\conference{\vspace{-4mm}}\arxiv{\vspace{-4mm}}
\caption{Regions and population distributions at 12:00. 
Blue and green squares represent Shinjuku and Akihabara, respectively. 
These areas are crowded in both the SNS-based people flow data and location data in our contest. 
}
\label{fig:regions}
\end{figure}

\smallskip
\noindent{\textbf{\colorB{Generation of Traces.}}}~~We generate synthetic traces for each team using our location synthesizer. 
Specifically, we use the SNS-based people flow data \cite{SNS_people_flow} (Tokyo) as training traces of our location synthesizer. 
We divide Tokyo equally into $32\times32$ regions ($1024$ regions in total) and assign region IDs sequentially from lower-left to upper-right. 
The size of each region is approximately $347$m (height) $\times$ $341$m (width). 
The left panel of Figure~\ref{fig:regions} shows the regions in our contest. 

Let $m \in \nats$ be the number of virtual users for each team. 
From the training traces, we generate synthetic traces of $m=2000$ virtual users for each team using our location synthesizer. 
For each virtual user, we generate traces from 8:00 to 18:00 for 40 days with a time interval of 30 minutes.
We use traces of the former 20 days as reference traces and the latter 20 days as original traces. 
\colorB{Here, we set the reference trace length to 20 days because the existing work assumes such a long reference trace, e.g., two years \cite{Gambs_JCSS14}, one month \cite{Mulder_WPES08}, or two to three weeks \cite{Pyrgelis_PoPETs17,Pyrgelis_NDSS18}.} 
Let $l_o, l_r \in \nats$ be the length of a training trace and reference trace, respectively. 
Then $l_o = l_r = 400$. 
Table~\ref{tab:loc_data} summarizes location data in our contest. 

\begin{table}[t]
\caption{Location Data in Our Contest}
\conference{\vspace{-4mm}}
\label{tab:loc_data}
\hbox to\hsize{\hfil
\arxiv{\footnotesize}
\begin{tabular}{c|c}
\hline 
Number of users	& $m=2000$\\
\hline 
Number of regions & $1024$ ($=32\times32$)\\
\hline 
Trace length 
&
$l_o=l_r=400$ 
(from $8:00$ to $18:00$ \\
& for $20$ days with 
$30$ minutes interval)\\
\hline
\end{tabular}
\hfil}
\end{table}

The middle and right panels of Figure~\ref{fig:regions} show population distributions at 12:00 in the SNS-based people flow data and synthetic traces in our contest, respectively. 
These panels show that the synthetic traces preserve the time-dependent population distribution. 

Let $O\tth$ and $R\tth$ be sets of original traces and reference traces for the $t$-th team $P_t$, respectively.

\subsection{Details of Our Contest}
\label{sub:details_contest}

Below we describe the details of our contest such as 
scenarios, \colorB{threat models,} 
contest flow, anonymization, privacy attacks, and utility and privacy scores.

\smallskip
\noindent{\textbf{Scenarios \colorB{and Threat Models}.}}~~\colorB{There are two possible scenarios in our contest. 
The first scenario is} 
\textit{geo-data analysis in 
a centralized model} \cite{DP}, where an LBS provider 
anonymizes location traces before providing them to a (possibly malicious) 
data analyst. 
\colorB{In this scenario, the data analyst can be an adversary.} 

\colorB{The second scenario} is 
\textit{LBS with an intermediate 
server} \cite{Bettini_SDM05,Gedik_TMC08}. 
In this scenario, each user sends her traces to a trusted intermediate server. 
Then the intermediate server anonymizes the traces of the users and sends them to a (possibly malicious) LBS provider.
Finally, each user receives \colorB{some services (i.e., personalized POI recommendation \cite{Cheng_IJCAI13,Feng_IJCAI15,Liu_VLDB17})} from the LBS provider through the intermediate server based on her anonymized traces. 
\colorB{Consider \textit{successive personalized POI recommendation} (or \textit{next POI recommendation}) \cite{Cheng_IJCAI13,Feng_IJCAI15} as an example. 
In this service, 
the LBS provider recommends 
a list of nearby POIs for each location visited by her. 
For example, if Alice visits a coffee shop in the morning, a university at noon, and a restaurant in the evening, then the LBS provider recommends POIs nearby the coffee shop, university, and restaurant. 
In this scenario, the LBS provider can be an (honest-but-curious) adversary.} 

In both of the scenarios, 
an adversary 
does not know the original traces and obtains the anonymized traces. 
\colorB{The adversary performs privacy attacks using reference traces of 20 days, which are separated from the original traces. 
This is consistent with the existing work that assumes such a long reference trace \cite{Gambs_JCSS14,Mulder_WPES08,Pyrgelis_PoPETs17,Pyrgelis_NDSS18}.} 

\begin{figure}[t]
\centering
\includegraphics[width=0.95\linewidth]{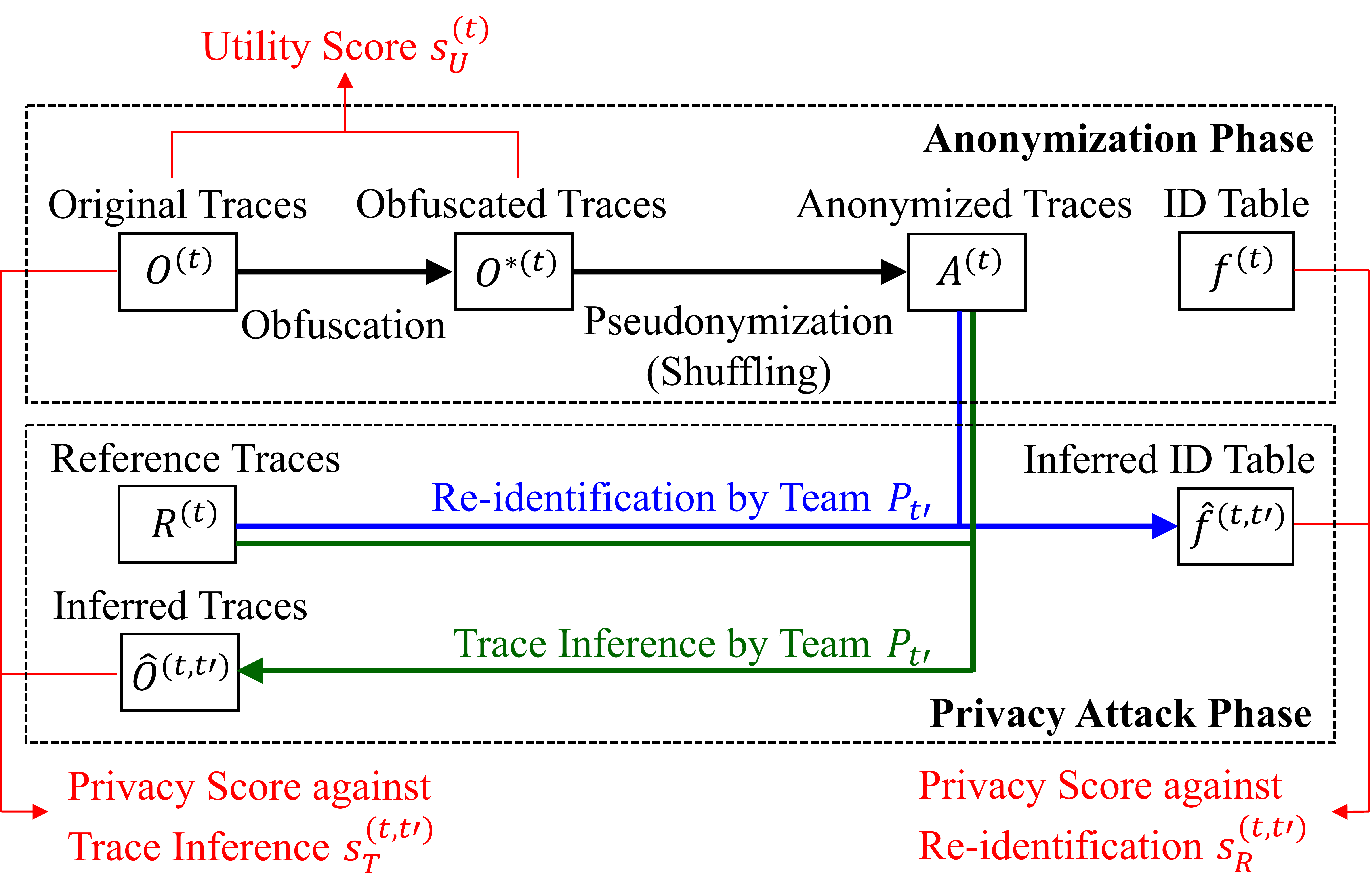}
\vspace{-3mm}
\caption{Contest flow for team $P_t$ ($t\in[\tau]$). Re-identification and trace inference are performed by another team $P_{t'}$ ($t' \ne t$). 
Utility and privacy scores are calculated by judge $Q$.}
\label{fig:contest_flow}
\end{figure}

\smallskip
\noindent{\textbf{Contest Flow.}}~~Figure~\ref{fig:contest_flow} shows our contest flow. 
In our contest, an organizer plays a role as a \textit{judge} who 
distributes traces for each team and evaluates 
privacy and utility scores of each team. 
\colorB{Note that the judging process (i.e., sending traces and calculating scores) can be automated.} 
Let $Q$ be a judge. 
Teams $P_1, \cdots, P_\tau$ and judge $Q$ participate in our contest.

In the anonymization phase, judge $Q$ distributes original traces $O\tth$ to each team $P_t$. 
Team $P_t$ obfuscates $O\tth$. 
Let 
$O\atth$ 
be obfuscated traces of team $P_t$. 
Team $P_t$ submits obfuscated traces $O\atth$ to 
$Q$. 
After receiving $O\atth$, 
$Q$ pseudonymizes $O\atth$ by randomly shuffling $m$ traces in $O\atth$ and then sequentially assigning pseudonyms. 
Consequently, 
$Q$ obtains 
anonymized traces 
$A\tth$ 
and an \textit{ID table} $f\tth$, which 
is a set of pairs 
between user IDs and pseudonyms. 
$Q$ keeps $f\tth$ secret.

Judge $Q$ calculates a utility score $s_U\tth \in [0,1]$ using original traces $O\tth$ and obfuscated traces $O\atth$. 
Then $Q$ compares $s_U\tth$ with a threshold $s_{req} \in [0,1]$ that is publicly available. 
If $s_U\tth \geq s_{req}$, then 
$Q$ regards anonymized traces $A\tth$ as \textit{valid} (otherwise, \textit{invalid}). 
Note that $s_U\tth$ is calculated from $O\tth$ and $O\atth$. 
Thus, 
$P_t$ can check whether 
$A\tth$ are valid before 
submitting $O\atth$ to $Q$. 

In the privacy attack phase, judge $Q$ distributes 
all reference traces and all \textit{valid} anonymized traces 
to all teams. 
Then each team attempts re-identification and trace inference against the valid anonymized traces of the other teams using their reference traces. 

Assume that 
team $P_{t'}$ ($t' \ne t$) 
attacks valid anonymized traces $A\tth$ of team $P_t$. 
As re-identification, team $P_{t'}$ infers user IDs corresponding to pseudonyms in $A\tth$ using reference traces $R\tth$. 
Team $P_{t'}$ 
creates 
an \textit{inferred ID table $\hf\ttp$}, which 
is a set of pairs 
between inferred user IDs and pseudonyms, and submits it to judge $Q$. 
As trace inference, team $P_{t'}$ infers all locations 
in original traces $O\tth$ from 
$A\tth$ using 
$R\tth$. 
Team $P_{t'}$ 
creates 
\textit{inferred traces} $\hO\ttp$, which include the inferred locations, and submits it to 
$Q$. 

Judge $Q$ calculates a re-identification privacy score $s_R\ttp \in [0,1]$ using ID table $f\tth$ and inferred ID table $\hf\ttp$. 
$Q$ also calculates a trace inference privacy score $s_T\ttp \in [0,1]$ using original traces $O\tth$ and inferred traces $\hO\ttp$.

Note that anonymized traces $A\tth$ get 
$\tau - 1$ attacks from the other teams. 
They also get attacks from some sample 
algorithms for 
re-identification and trace inference, which are described in Section~\ref{sec:pre_exp}. 
Judge $Q$ calculates privacy scores against all of these attacks and finds 
the minimum privacy score. 
Let $s_{R,min}\tth$ (resp.~$s_{T,min}\tth$) $\in [0,1]$ be the minimum 
re-identification (resp.~trace inference) privacy score of team $P_t$. 
$s_{R,min}\tth$ and $s_{T,min}\tth$ are final privacy scores of team $P_t$; i.e., we adopt a privacy score by the strongest attack. 

\smallskip
\noindent{\textbf{Pseudonymized Traces.}}~~In the privacy attack phase, judge $Q$ also distributes \textit{pseudonymized traces} prepared by $Q$ and makes each team attack these traces. 
The purpose of this is to 
compare the 
privacy 
of each 
team's 
anonymized traces 
with that of the pseudonymized traces; 
i.e., they play a role as a benchmark. 

The pseudonymized traces are generated as follows. 
Judge 
$Q$ generates reference traces $R^{(\tau+1)}$ and original traces $O^{(\tau+1)}$ for team $P^{(\tau+1)}$ (who does not participate in the contest). 
Then $Q$ 
makes 
anonymized traces 
$A^{(\tau+1)}$ 
by only pseudonymization. 
Finally, $Q$ distributes $A^{(\tau+1)}$ and $R^{(\tau+1)}$, and each team $P_t$ attempts 
privacy attacks against 
$A^{(\tau+1)}$ using $R^{(\tau+1)}$. 

\smallskip
\noindent{\textbf{\colorB{Sample Traces.}}}~~\colorB{Judge $Q$ also generates reference and original traces of two teams $P^{(\tau+2)}$ and $P^{(\tau+3)}$ (who do not participate in the contest) as sample traces. $Q$ distributes the sample traces to all teams in the anonymization phase.  
The purpose of distributing sample traces is to allow each team to tune parameters in her anonymization and privacy attack algorithms.} 

\smallskip
\noindent{\textbf{Awards.}}~~As described in Section~\ref{sub:overview_contest}, awards are important to 
make 
both defense and attack 
complete together. 
We give the \textit{best anonymization award} to a team $P_t$ who achieves the 
highest 
trace inference privacy score $s_{T,min}\tth$ 
among all teams. 
We also give 
the \textit{best re-identification (resp.~trace inference) award} to a team $P_{t'}$ whose 
$\sum_{t=1}^{\tau+1} s_R\ttp$ 
(resp.~$\sum_{t=1}^{\tau+1} s_T\ttp$) is the lowest among all teams.

\smallskip
\noindent{\textbf{\colorB{Fairness.}}}~~\colorB{It should be noted that the diversity of location traces may raise a fairness issue. 
For example, even if 
all teams apply the same anonymization and attack algorithms, 
privacy scores can be different 
among the teams.} 

\colorB{In Appendix~\ref{app:fairness_contest}, we evaluate the fairness of our contest. 
Specifically, we evaluate the variance of privacy scores against the same attack algorithm and show that it is small. 
For example, the standard deviation of re-identification privacy scores is about $0.01$ or less, which is much smaller than the difference between the best privacy score ($=0.79$) and the second-best privacy score ($=0.67$). 
See Appendix~\ref{app:fairness_contest} for more details.}

\smallskip
\noindent{\textbf{Anonymization.}}~~In the anonymization phase, team $P_t$ obfuscates 
its original traces $O\tth$. 
In our contest, we allow four types of processing for each location:
\begin{enumerate}
    \item \textbf{No Obfuscation}: 
    Output the original location as is; e.g., $x_1 \rightarrow x_1$.
    \item \textbf{Perturbation (Adding Noise):} 
    Replace the original location with another location; e.g., $x_1 \rightarrow x_2$. 
    \item \textbf{Generalization:} Replace the original location with a set of multiple locations; 
    e.g., $x_1 \rightarrow \{x_1, x_3\}$, $x_1 \rightarrow \{x_2, x_3, x_5\}$. 
    Note that the original location may not be included in the set.
    \item \textbf{Deletion:} Replace the original location with an empty set $\emptyset$ representing deletion; e.g., $x_1 \rightarrow \emptyset$. 
\end{enumerate}
Let $\calY$ be a finite set of outputs after applying one of the four types of processing to a location. 
Then $\calY$ is represented as a power set of $\calX$; i.e., 
$\calY = 2^\calX$. 
In other words, we accept \textit{all possible operations} on each location. 

Then 
judge $Q$ pseudonymizes 
obfuscated traces $O\atth$. 
Specifically, judge $Q$ randomly permutes $1, \cdots, m$ and sequentially assign pseudonyms $m+1, \cdots, 2m$. 

\begin{figure}[t]
\centering
\includegraphics[width=0.99\linewidth]{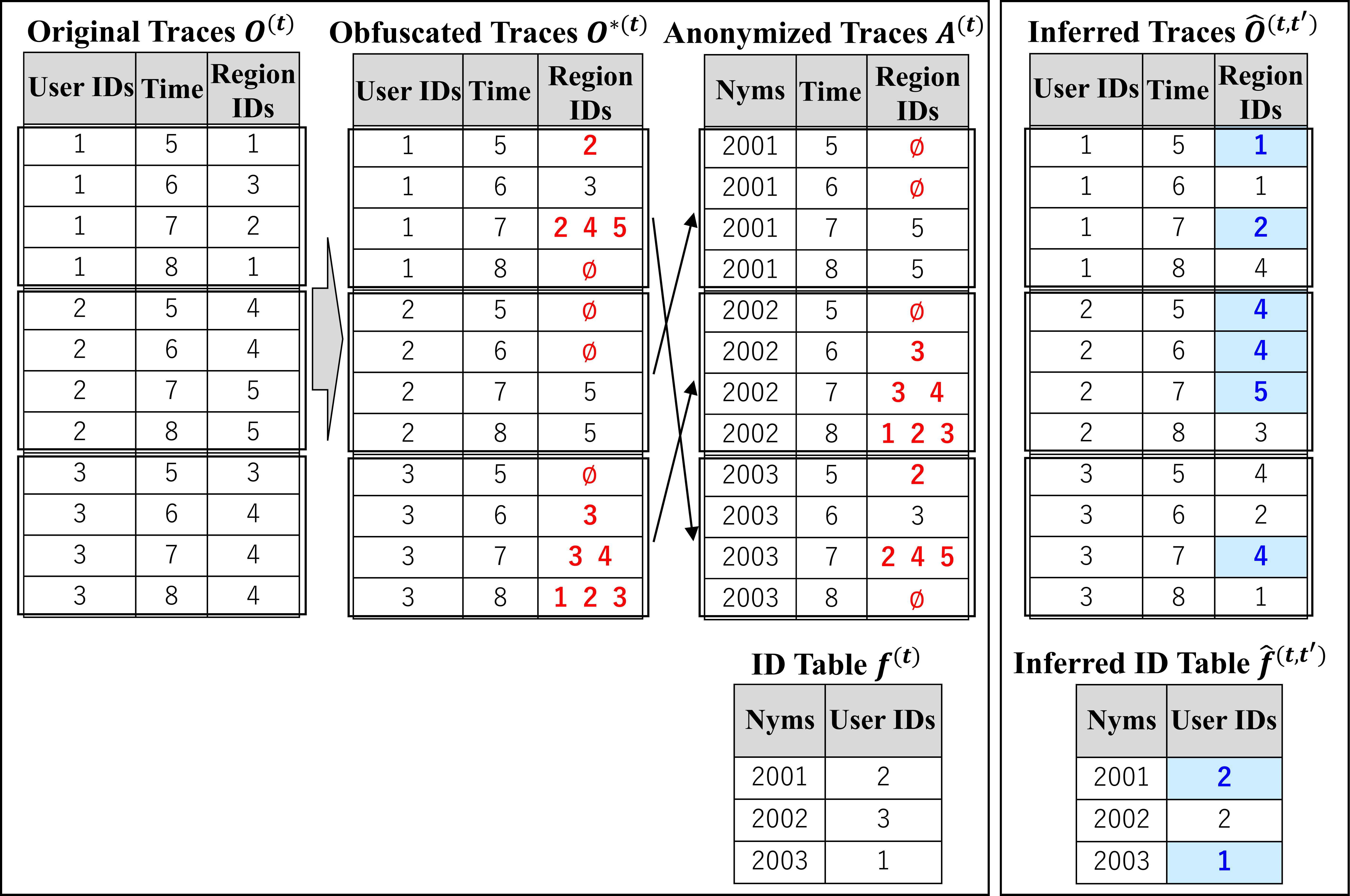}
\conference{\vspace{-3mm}}\arxiv{\vspace{-3mm}}
\caption{Example of anonymization and privacy attacks. 
Obfuscated locations are marked with bold red font. 
``2 4 5'' represents generalized locations $\{x_2, x_4, x_5\}$. 
``$\emptyset$'' represents deletion. 
Correct user/region IDs are marked with bold blue font on a light blue background.}
\label{fig:anonymization}
\end{figure}

The left panel of 
Figure~\ref{fig:anonymization} shows an example of anonymization, 
where user IDs and region IDs are subscripts of users and regions, respectively. 
In this example, 
pseudonyms $2001$, $2002$, and $2003$ 
correspond to user IDs $2$, $3$, and $1$, respectively; i.e., 
$f\tth = \{(2001,2), (2002,3), (2003,1)\}$. 

\smallskip
\noindent{\textbf{Privacy Attack.}}~~In the privacy attack phase, team $P_{t'}$ ($t' \ne t$) attempts 
privacy attacks 
against (valid) anonymized traces $A\tth$ of team $P_t$ using reference traces $R\tth$. 
Specifically, team $P_{t'}$ creates an inferred ID table $\hf\ttp$ and inferred traces $\hO\ttp$ for re-identification and trace inference, respectively. 
Here we allow 
$P_{t'}$ to identify multiple pseudonyms in $A\tth$ as the same user ID. 

The right panel of 
Figure~\ref{fig:anonymization} 
shows an example of 
privacy attacks. 
In this example, 
$\hf\ttp = \{(2001,2), (2002,2), \allowbreak (2003,1)\}$. 

\smallskip
\noindent{\textbf{Utility Score.}}~~Anonymized traces $A\tth$ are useful for 
geo-data analysis, 
e.g., 
mining popular POIs \cite{Zheng_WWW09}, 
auto-tagging POI categories \cite{Do_TMC13,Ye_KDD11}, 
and modeling human mobility patterns \cite{Liu_CIKM13,Song_TMC06}. 
They are also useful for LBS 
in the intermediate server model (as described in Section~\ref{sub:details_contest} ``Scenarios and Threat Models''). 
\colorB{For example, in the successive personalized POI recommendation \cite{Cheng_IJCAI13,Feng_IJCAI15}, it is important to preserve rough information about each location in the original traces. 
Thus, a location synthesizer that preserves only statistical information about the original traces is not useful as an anonymization method in the latter scenario.} 
To accommodate a variety of purposes, we adopt a versatile utility score $s_U\tth \in [0,1]$. 

Specifically, \colorB{for both geo-data analysis and LBS,} 
it would be natural to consider that the utility degrades as the distance between an original location and a noisy location becomes larger. 
The utility would be completely lost when the distance exceeds a certain level or when the original location is deleted. 

Taking this into account, we define the utility score $s_U\tth$. 
Our utility score is similar to the service quality loss (SQL) \cite{Andres_CCS13,Bordenabe_CCS14,Shokri_CCS12} for perturbation 
in that the utility is measured by the expected Euclidean distance between original locations and obfuscated locations. 
Our utility score differs from the SQL in two ways: 
(i) we deal with perturbation, generalization, and deletion; (ii) we assume the utility is completely lost when the distance exceeds a certain level or the original location is deleted. 

Formally, let $d: \calX \times \calX \rightarrow \nngreals$ be a distance function that takes two locations $x_i, x_j \in \calX$ as input and outputs their Euclidean distance $d(x_i, x_j) \in \nngreals$. 
Since the location data are regions 
in our contest, we define $d(x_i, x_j)$ as the Euclidean distance between center points of region $x_i$ and $x_j$. 
For example, $d(x_1, x_2) = 341$m and $d(x_1, x_{34}) = \sqrt{347^2 + 341^2} = 487$m in our contest, as the size of each region is $347$m (height) $\times$ $341$m (width). 

We calculate the 
Euclidean distance between each location in $O\tth$ and the corresponding location(s) in $O\atth$. 
For $i \in [m]$ and $j \in [l_o]$, 
let $\alpha_{i,j}\tth \in \nngreals$ be the 
Euclidean distance between the $j$-th locations in the original and obfuscated traces for user $v_i\tth$. 
$\alpha_{i,j}\tth$ takes 
the \textit{average} Euclidean distance for generalization, and $\infty$ for deletion. 
For example, 
$\alpha_{1,1}\tth = d(x_1, x_2)$, 
$\alpha_{1,2}\tth = d(x_3, x_3) = 0$, 
$\alpha_{1,3}\tth = \frac{d(x_2, x_2)+d(x_2, x_4)+d(x_2, x_5)}{3}$, and 
$\alpha_{1,4}\tth = \infty$ in Figure~\ref{fig:anonymization}. 

\begin{figure}
\centering
\includegraphics[width=0.65\linewidth]{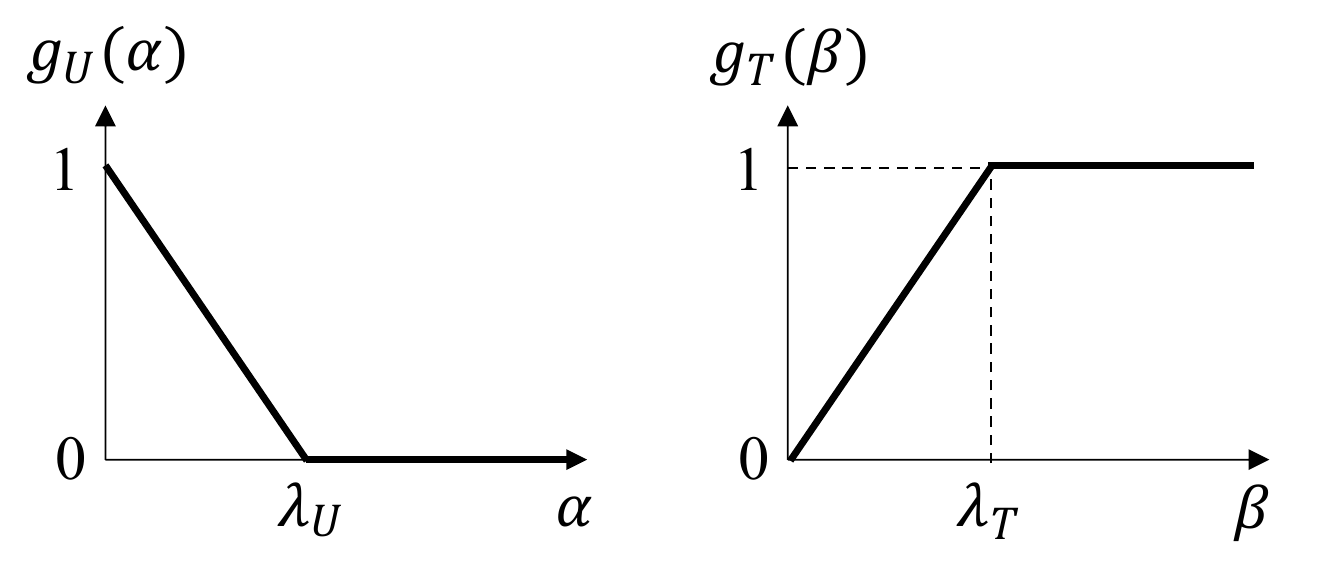}
\vspace{-3mm}
\caption{Piecewise linear functions $g_U$ and $g_T$.}
\label{fig:piecewise}
\end{figure}

Finally, we use a piecewise linear function $g_U$ shown in the left of Figure~\ref{fig:piecewise} to transform each $\alpha_{i,j}\tth$ into a score value from $0$ to $1$ (higher is better). 
Then we calculate the utility score $s_U\tth$ by taking the average of $m l_o$ scores. 

Specifically, let $g_U: \nngreals \rightarrow [0,1]$ be a function that takes $\alpha \in \nngreals$ as input and outputs the following score: 
\begin{align*}
    g_U(\alpha) = 
    \begin{cases}
    1 - \frac{\alpha}{\lambda_U} &   \text{(if $\alpha < \lambda_U$)}\\
    0               &   \text{(if $\alpha \geq \lambda_U$)},
    \end{cases}
\end{align*}
where $\lambda_U \in \nngreals$ is a 
threshold. 
Using this function, we calculate the utility score $s_U\tth$ 
as follows: 
\begin{align*}
\textstyle{s_U\tth = \frac{1}{m l_o} \sum_{i=1}^m \sum_{j=1}^{l_o} g_U(\alpha_{i,j}\tth).}
\end{align*}
For example, if we do not obfuscate any location, 
then $s_U\tth = 1$. 
If we delete all locations or the Euclidean distance exceeds $\lambda_U$ for all locations, then $s_U\tth = 0$. 

In our contest, we set the threshold $\lambda_U$ 
to $\lambda_U=2$km and the threshold $s_{req}$ of the utility score (for determining whether or not anonymized traces are valid) 
to $s_{req}=0.7$. 
In 
Section~\ref{sub:utility_contest}, 
we also show that 
valid anonymized traces have high utility 
for a variety of purposes such as POI recommendation and geo-data analysis.

\smallskip
\noindent{\textbf{Re-identification Privacy Score.}}~~A 
re-identification privacy score $s_R\ttp \in [0,1]$ 
is calculated by comparing ID table $f\tth$ with inferred ID table $\hf\ttp$. 

In our contest, we calculate 
$s_R\ttp$ 
based on the \textit{re-identification rate} \cite{Gambs_JCSS14,Mulder_WPES08,Murakami_PoPETs21}, 
a proportion of correctly identified pseudonyms. 
Specifically, we calculate $s_R\ttp$ by subtracting the re-identification rate from 1 (higher is better). 
For example, 
$s_R\ttp = 1 - \frac{2}{3} = \frac{1}{3}$ 
in Figure~\ref{fig:anonymization}. 

\smallskip
\noindent{\textbf{Trace Inference Privacy Score.}}~~A 
trace inference privacy score $s_T\ttp \in [0,1]$ 
is calculated by comparing original traces $O\tth$ with inferred traces $\hO\ttp$. 

\colorB{Since Shokri \textit{et al.} \cite{Shokri_SP11} showed that incorrectness determines the privacy of users, the adversary's \textit{expected error} has been widely used as a location privacy metric. 
The expected error is an average distance (e.g., the Euclidean distance \cite{Shokri_CCS12}) between original locations and inferred locations.} 

Formally, for $i \in [m]$ and $j \in [l_o]$, let $\beta_{i,j}\ttp \in \nngreals$ be the Euclidean distance between the $j$-th locations in the original and inferred traces for user $v_i\tth$. 
For example, 
$\beta_{1,1}\ttp = d(x_1, x_1)$, 
$\beta_{1,2}\ttp = d(x_3, x_1)$, 
$\beta_{1,3}\ttp = d(x_2, x_2)$, and 
$\beta_{1,4}\ttp = d(x_1, x_4)$ in Figure~\ref{fig:anonymization}. 
\colorB{Then, the expected error with the Euclidean metric is given by:} 
\begin{align}
\colorB{\textstyle{\frac{1}{m l_o} \sum_{i=1}^m \sum_{j=1}^{l_o} \beta_{i,j}\ttp.}}
\label{eq:expected_error}
\end{align}

In our contest, 
\colorB{we use the expected error with two modifications. 
First, we want all utility and privacy scores to be between 0 and 1 (higher is better) so that they are easy to understand for all teams. Thus, 
we transform the Euclidean distance into a score value from $0$ to $1$.} 
Specifically, 
we assume that the adversary completely fails to infer the original location when the distance exceeds a certain level. 
In other words, we use a piecewise linear function $g_T$ in the right of Figure~\ref{fig:piecewise} to transform each $\beta_{i,j}\ttp$ into a score value from $0$ to $1$ (higher is better). 
Formally, 
let $g_T: \nngreals \rightarrow [0,1]$ be a function that takes $\beta \in \nngreals$ as input and outputs the following score: 
\begin{align*}
    g_T(\beta) = 
    \begin{cases}
    \frac{\beta}{\lambda_T} &   \text{(if $\beta < \lambda_T$)}\\
    1               &   \text{(if $\beta \geq \lambda_T$)},
    \end{cases}
\end{align*}
where $\lambda_T \in \nngreals$ is a threshold. 
In our contest, we set $\lambda_T = 2$km. 
By using 
$g_T$, we obtain $m l_o$ scores.

\colorB{Second, 
we consider regions that include hospitals (referred to as \textit{hospital regions}) to be especially sensitive.} 
There are 
$37$ hospital regions 
in the SNS-based people flow data \cite{SNS_people_flow}. 
Since sensitive locations need to be carefully handled, 
we calculate the privacy score $s_T\ttp$ by taking a \textit{weighted} average of $m l_o$ scores, where 
\colorB{we set a weight value to $10$ for hospital regions and $1$ for the others}. 

Thus, 
we calculate the privacy score $s_T\ttp$ as: 
\begin{align}
\textstyle{s_T\ttp = \frac{\sum_{i=1}^m \sum_{j=1}^{l_o} w_{i,j} g_T(\beta_{i,j}\ttp)}{\sum_{i=1}^m \sum_{j=1}^{l_o} w_{i,j}},}
\label{eq:modified_expected_error}
\end{align}
where $w_{i,j} \in \{1,10\}$ is a weight variable that takes $10$ if the $j$-th location in the original trace of user $v_i\tth$ is a hospital region, and $1$ otherwise. 
If the inferred traces are perfectly correct 
($O\tth = \hO\ttp$), then $s_T\ttp = 0$. 

\colorR{We set a hospital weight to $10$ because a too large value results in a low correlation between our privacy score and the expected error, as shown in Section~\ref{sub:other_metrics}. 
In other words, if we choose a too large hospital weight, the best anonymization algorithm loses its versatility -- it may not be useful when the expected error is used as a privacy metric. 
Our privacy score with hospital weight $=10$ carefully handles sensitive regions (as hospital weight $\gg 1$) and is yet highly correlated with the expected error.}

\smallskip
\noindent{\textbf{\colorR{Remark.}}}~~\colorR{Because our utility/privacy scores are based on the existing metrics (i.e., SQL \cite{Andres_CCS13,Bordenabe_CCS14,Shokri_CCS12}, re-identification rate \cite{Gambs_JCSS14,Mulder_WPES08,Murakami_PoPETs21}, and the expected error \cite{Shokri_SP11}), they are also useful for 
evaluating anonymization techniques in research papers. 
One difference between our contest and academic research is that both defense and attack compete together (i.e., each team attacks other teams) in our contest. 
Such evaluation might be difficult for academic research.} 

\section{Preliminary Experiments}
\label{sec:pre_exp}
\colorB{Prior to our contest, we conducted preliminary experiments. 
The main purpose of the preliminary experiments is to show that the cheating anonymization \cite{Kikuchi_AINA16}, which is perfectly secure against re-identification as described in Section~\ref{sec:intro}, is not secure against trace inference. 
This result serves as strong evidence that trace inference should be added as an additional risk in our contest. 
Another purpose is to make all teams understand how to perform obfuscation, re-identification, and trace inference. 
To this end, we implemented some sample algorithms and released all the sample algorithms and the experimental results to all teams before the contest. 
Section~\ref{sub:pre_setup} explains our experimental set-up. 
Section~\ref{sub:pre_res} reports our experimental results.} 

\subsection{Experimental Set-up}
\label{sub:pre_setup}
\noindent{\textbf{Dataset.}}~~We 
used the SNS-based people flow data \cite{SNS_people_flow} (Osaka). 
We divided Osaka 
into $32 \times 32$ regions 
($1024$ regions in total), 
and 
extracted training traces from 8:00 to 18:00 for $4071$ users. 
We trained our location synthesizer 
using the training traces.

Then we generated reference traces 
$R^{(0)}$ 
and original traces 
$O^{(0)}$ 
for $m=2000$ virtual users in one team 
(whose team number is $t=0$) 
using our synthesizer. 
Each trace includes locations from 8:00 to 18:00 for $20$ days with time interval of 30 minutes ($l_o = l_r = 400)$. 

\smallskip
\noindent{\textbf{Sample Algorithms.}}~~We implemented some sample algorithms for obfuscation, re-identification attacks, and trace inference attacks. 
We anonymized the original traces $O^{(0)}$ by using each sample obfuscation algorithm. 
Then we performed privacy attacks by using each sample re-identification or trace inference algorithm. 

For obfuscation, we implemented the following algorithms:
\begin{itemize}
    \item 
    \textano{No Obfuscation}: 
    Output the original location as is. 
    In other words, we perform only pseudonymization.
    \item \textano{MRLH($\mu_x,\mu_y,\lambda$)}: 
    Merging regions and location hiding in \cite{Shokri_SP11}. 
    It generalizes each region in the original trace by dropping lower $\mu_x$ (resp.~$\mu_y$) bits of the $x$ (resp.~$y$) coordinate expressed as a binary sequence and deletes the region with probability $\lambda$. 
    For example, the $x$ (resp.~$y$) coordinate of $x_2$ is $00001$ (resp.~$00000$) in Figure~\ref{fig:regions}. Thus, given input $x_2$, MRLH($1,1,0.8$) outputs $\{x_1, x_2, x_{33}, x_{34}\}$ with probability $0.2$ and deletes $x_2$ with probability $0.8$.
    \item \textano{RR($\epsilon$)}: 
    The 
    $\kappa$-ary randomized response in \cite{Kairouz_ICML16}, where $\kappa$ is the size of input domain ($\kappa=1024$ in our experiments).  
    It outputs the original region with probability 
    $\frac{e^\epsilon}{\kappa-1+e^\epsilon}$, 
    and outputs another region at random with the remaining probability. 
    It provides $\epsilon$-DP for each region. 
    \item \textano{PL($l, r$)}: 
    The planar Laplace mechanism \cite{Andres_CCS13}. 
    It perturbs each region in the original trace according to the planar Laplacian distribution so that it provides $l$-DP within $r$ km. 
    This privacy property is known as $\epsilon$-geo-indistinguishability \cite{Andres_CCS13}, where $\epsilon=l/r$.
    \item \textano{Cheat($p$)}: 
    The cheating anonymization 
    \cite{Kikuchi_AINA16}. 
    It selects the first $p$ ($0 \leq p \leq 1$) of all users 
    as a subset of users, 
    and randomly shuffles the whole traces within the subset (as in Figure~\ref{fig:cheat}). 
    Note that this is \textit{excessive location obfuscation} rather than pseudonymization.
\end{itemize}

\colorB{For privacy attacks, we developed two sample algorithms for re-identification (\textano{VisitProb-R}, \textano{HomeProb-R}) and two algorithms for trace inference (\textano{VisitProb-T}, \textano{HomeProb-T}).
All of them are based on a \textit{visit probability vector}, which 
comprises 
the visit probability for each region. 
We calculate the visit probability vector for each virtual user based on reference traces. 
Then we perform re-identification or trace inference for each anonymized trace 
using the visit probability vectors. 
We published all the sample algorithms 
as open-source software \cite{PWSCup2019}.} 

Below, we explain each attack algorithm. 
\colorR{We also show examples of visit probability vectors, \textano{VisitProb-R}, and \textano{VisitProb-T} in Appendix~\ref{app:examples_sample}.}

\smallskip
\noindent{\textbf{VisitProb-R.}}~~\textano{VisitProb-R} 
first trains a visit-probability vector 
for each user 
from 
reference traces. 
For an element with zero probability, it assigns a very small positive value $\delta$ ($=10^{-8}$) to guarantee that the likelihood never becomes zero. 

Then \textano{VisitProb-R} re-identifies each trace as follows. 
It computes the likelihood (the product of the likelihood for each region) for each user. 
For generalized regions, it averages the likelihood over generalized regions. 
For deletion, it does not update the likelihood. 
After computing the likelihood for each user, it outputs a user ID with the highest likelihood as an identification result. 

\smallskip
\noindent{\textbf{HomeProb-R.}}~~\textano{HomeProb-R} re-identifies traces based on the fact that a user tends to be at her home region between 8:00 and 9:00 (see Appendix~\ref{app:home_regions} for details). 
Specifically, it modifies \textano{VisitProb-R} 
to use only regions between 8:00 and 9:00. 

\smallskip
\noindent{\textbf{VisitProb-T.}}~~\textano{VisitProb-T} first re-identifies traces using \textano{VisitProb-R}. 
Here, it does not choose an already re-identified user to avoid duplication of user IDs. 
Then it de-obfuscates the original regions for each re-identified trace. 
For perturbation, it outputs the noisy location as is. 
For generalization, it randomly chooses a region from generalized regions. 
For deletion, it randomly chooses a region from all regions. 

\smallskip
\noindent{\textbf{HomeProb-T.}}~~\textano{HomeProb-T} modifies 
\textano{VisitProb-T} to use \textano{HomeProb-R} for re-identification.

\subsection{Experimental Results}
\label{sub:pre_res}

\begin{figure}
\centering
\includegraphics[width=0.92\linewidth]{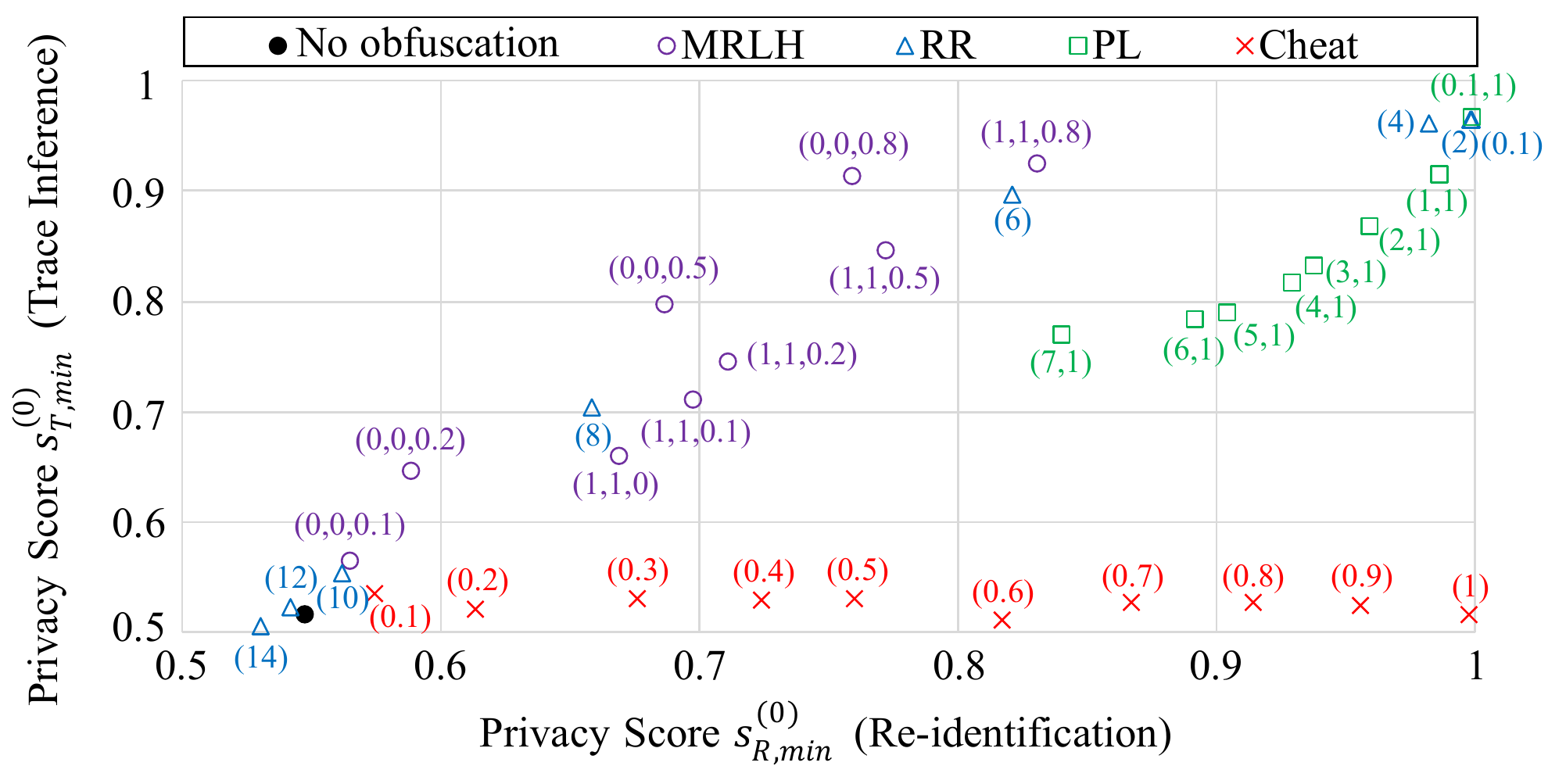}
\vspace{-4mm}
\caption{Privacy scores of the sample algorithms (higher is better). The numbers in parentheses represent the parameters in the sample algorithms.} 
\label{fig:pre_exp}
\end{figure}

\noindent{\textbf{Results.}}~~For each sample obfuscation algorithm, we calculated the minimum 
re-identification (resp.~trace inference) 
privacy score $s_{R,min}^{(0)}$ (resp.~$s_{T,min}^{(0)}$) 
over the sample attack algorithms. 
Figure~\ref{fig:pre_exp} shows the results. 

Overall, there is a positive correlation between 
the re-identification privacy score $s_{R,min}^{(0)}$ and the trace inference privacy score $s_{T,min}^{(0)}$. 
However, there is 
a clear 
exception -- \textit{cheating anonymization}. 
In cheating anonymization, $s_{R,min}^{(0)}$ increases with increase in the parameter $p$. 
When $p=1$ (i.e., when we shuffle all users), 
$s_{R,min}^{(0)}$ is almost $1$. 
In other words, the re-identification rate is almost $0$ for \textano{Cheat($1$)}. 
This is because the adversary cannot find which permutation is correct, as described in Section~\ref{sec:intro}. 
Thus, the adversary 
cannot re-identify traces with higher accuracy than a random guess ($=1/2000$).
However, $s_{T,min}^{(0)}$ does \textit{not} increase with increase in $p$, and $s_{T,min}^{(0)}$ of 
\textano{Cheat($1$)} 
is almost the same as that of 
\textano{No Obfuscation}. 
This means that the adversary can recover the original traces from anonymized traces \textit{without} accurately re-identifying them.

We can explain why this occurs as follows. 
Suppose that the adversary has reference traces highly correlated with the original traces 
in the example of Figure~\ref{fig:cheat}. 
First, this adversary would re-identify $10001$ as $v_2$, which is incorrect. 
Then, the adversary may recover the trace of $v_2$ as $x_1 \rightarrow x_1 \rightarrow x_1 \rightarrow x_2$ because they are included in the anonymized trace of $10001$. 
This is perfectly correct.

This example explains the intuition that the cheating anonymization is insecure -- 
the adversary 
can easily recover the original traces from the anonymized traces, even if she cannot accurately re-identify them. 
Figure~\ref{fig:pre_exp} clearly shows that 
the re-identification alone is insufficient as a privacy risk for 
the cheating anonymization. 

\smallskip
\noindent{\textbf{\colorB{Take Aways.}}}~~\colorB{In summary, we should avoid using re-identification alone as a privacy metric when organizing a contest. 
Otherwise, there is no guarantee that a winning team's algorithm, which achieves the highest re-identification privacy score, protects user privacy. 
As described in Section~\ref{sec:intro}, 
$k$-anonymity is also vulnerable to attribute (location) inference \cite{Machanavajjhala_ICDE06}. 
However, our experimental results provide stronger evidence in that there is an algorithm that is \textit{perfectly} secure against re-identification and is not secure against trace inference. 
To make the contest meaningful, we should add trace inference as a risk.}

\section{Contest Results and Analysis}
\label{sec:results}
\colorB{We released all the sample algorithms and the results of our preliminary experiments to all teams before the contest.} 
Then we held our contest 
to answer the second question RQ2 in Section~\ref{sub:purpose_approach}. 
\colorB{Section~\ref{sub:contest_res} reports our contest results. 
Sections~\ref{sub:best_algorithm} and ~\ref{sub:ri_ti_teams} explain the best anonymization and privacy attack algorithms that won first place in our contest. 
Section~\ref{sub:other_metrics} analyzes the relationship between the expected error in \cite{Shokri_SP11} and our privacy scores with various weight values. 
Finally, Section~\ref{sub:utility_contest} shows that the anonymized traces in our contest are useful for various applications.}

\subsection{\colorB{Contest Results}}
\label{sub:contest_res}
\noindent{\textbf{Number of Teams.}}~~A total of $21$ teams participated in our contest ($\tau=21$). 
In the anonymization phase, $18$ teams submitted their obfuscated traces so that the anonymized traces were secure against trace inference. 
We set the threshold $s_{req}$ of the utility score to 
$s_{req}=0.7$, as described in Section~\ref{sub:details_contest}. 
The anonymized traces of $17$ (out of $18$) teams were valid. 

In the privacy attack phase, each team attempted re-identification and trace inference against the valid anonymized traces of the other teams 
and pseudonymized traces prepared by the organizer. 
Then we evaluated the minimum privacy 
scores $s_{R,min}\tth$ and $s_{T,min}\tth$ 
for 
the anonymized traces of the $17$ teams and the pseudonymized traces. 

\begin{figure}
\centering
\includegraphics[width=0.9\linewidth]{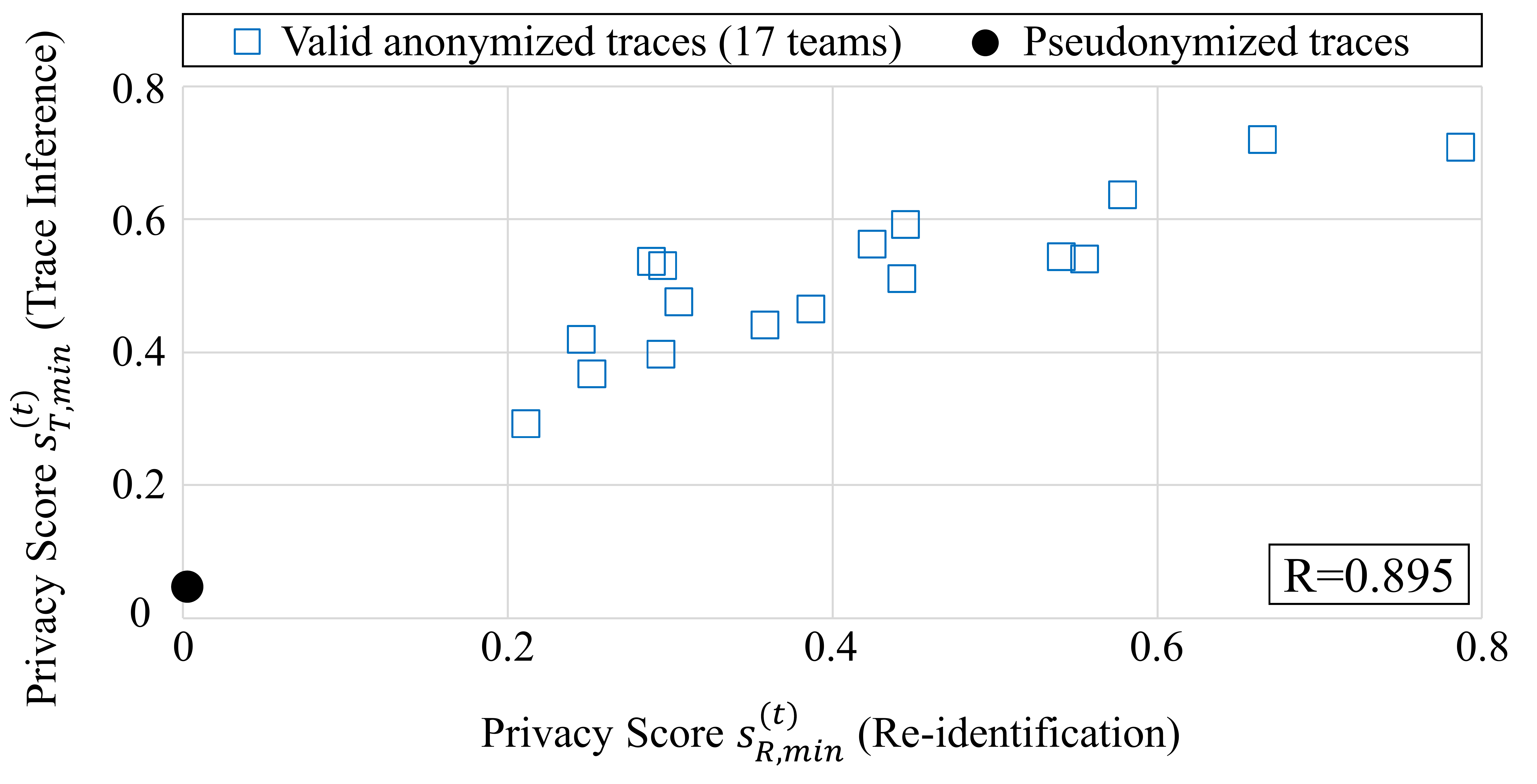}
\vspace{-4mm}
\caption{Privacy scores of the valid anonymized traces of the $17$ teams and the pseudonymized traces (higher is better). 
R is the correlation coefficient between $s_{R,min}\tth$ and $s_{T,min}\tth$.
} 
\label{fig:contest_res}
\end{figure}

\smallskip
\noindent{\textbf{Results.}}~~Figure~\ref{fig:contest_res} shows the results. 
It shows that there is a strong correlation between 
the re-identification privacy score $s_{R,min}\tth$ and the trace inference privacy score $s_{T,min}\tth$ 
(the correlation coefficient is $0.895$). 
We will discuss the reason for this 
\colorB{at the end of Section~\ref{sub:contest_res}}. 

Figure~\ref{fig:contest_res} also shows 
that the privacy of the pseudonymized traces is completely violated in terms of both re-identification and trace inference. 
This means that attacks by the teams are much stronger than the sample attacks.

\colorR{After the contest, all teams presented their algorithms in person. 
Thus, they learned which algorithm won and why. 
They also learned that pseudonymization is insufficient by violating pseudonymized traces by themselves. 
All of them play an educational role.} 

We also published the submitted files by all the teams 
\cite{PWSCup2019}. 

\smallskip
\noindent{\textbf{\colorB{Answer to RQ2 in Section~\ref{sub:purpose_approach}.}}}~~Figure~\ref{fig:contest_res} shows that there is a strong correlation between 
two privacy scores $s_{R,min}\tth$ and $s_{T,min}\tth$. 

The reason for this can be explained as follows. 
In our contest, we gave the best anonymization award to a team that achieved the highest privacy score $s_{T,min}\tth$ against trace inference. 
Thus, no team used the cheating anonymization that was not effective for trace inference\footnote{Another reason for not using the cheating anonymization 
is that it has a non-negligible impact on utility for our utility measure that performs a comparison trace by trace. However, even if we use a utility measure 
in which the cheating anonymization does not have any impact on utility (e.g., utility of aggregate information \cite{Pyrgelis_NDSS18}), this anonymization is still not effective for trace inference. 
Therefore, our conclusion here would not be changed.}. 
Consequently, 
each team had to re-identify traces and then de-obfuscate traces to recover the original traces. 
In other words, 
it was difficult to accurately recover the original traces without accurately identifying them. 
Moreover, all the traces were appropriately pseudonymized (randomly shuffled) by the organizer in our contest. 
Thus, re-identification was also difficult for traces that were well obfuscated. 
In this case, the accuracy of re-identification is closely related to the accuracy of trace inference. 
The team that won the best anonymization award also obfuscated traces so that re-identification was difficult 
(see Section~\ref{sub:best_algorithm} for details). 

In summary, under the presence of appropriate pseudonymization, the answer to RQ2 in Section~\ref{sub:purpose_approach} was yes in our contest.

\subsection{Best Anonymization Algorithm}
\label{sub:best_algorithm}

Below, we briefly explain the best anonymization algorithm that 
achieved the highest 
trace inference 
privacy score $s_{T,min}\tth$\footnote{\colorB{Note that we only report the trace inference challenge in this paper (see footnote 2). The best anonymization algorithm in the re-identification challenge was different.}}. 
The source code is also published in \cite{PWSCup2019}\footnote{\colorB{We obtained permission from the best teams to publish their algorithms on this paper and their source code on the website \cite{PWSCup2019}.}}.

The best anonymization team is from a company. 
The age range is 20s to 30s. 
The team members have participated in past PWS Cups. 
They are also a certified business operator of anonymization. 

\smallskip
\noindent{\textbf{\colorB{Algorithm.}}}~~\colorB{In a nutshell, the best anonymization algorithm obfuscates traces using $k$-means clustering so that re-identification is difficult within each cluster.} 

Specifically, the best algorithm consists of three steps: 
(i) clustering users based on visit count vectors, 
(ii) adding noise to regions so that visit count vectors are as similar as possible within the same cluster, 
and (iii) 
replacing each hospital region with another nearby hospital region. 
The first and second steps aim at preventing re-identification based on visit probability vectors. 
In addition, the amount of noise in step (ii) is small because the visit count vectors are similar within the cluster from the beginning. 
The third step aims at preventing the inference of sensitive hospital regions. 
The amount of noise in step (iii) is also small because the selected hospital region is close to the original hospital region.

\begin{figure}[t]
\centering
\includegraphics[width=0.85\linewidth]{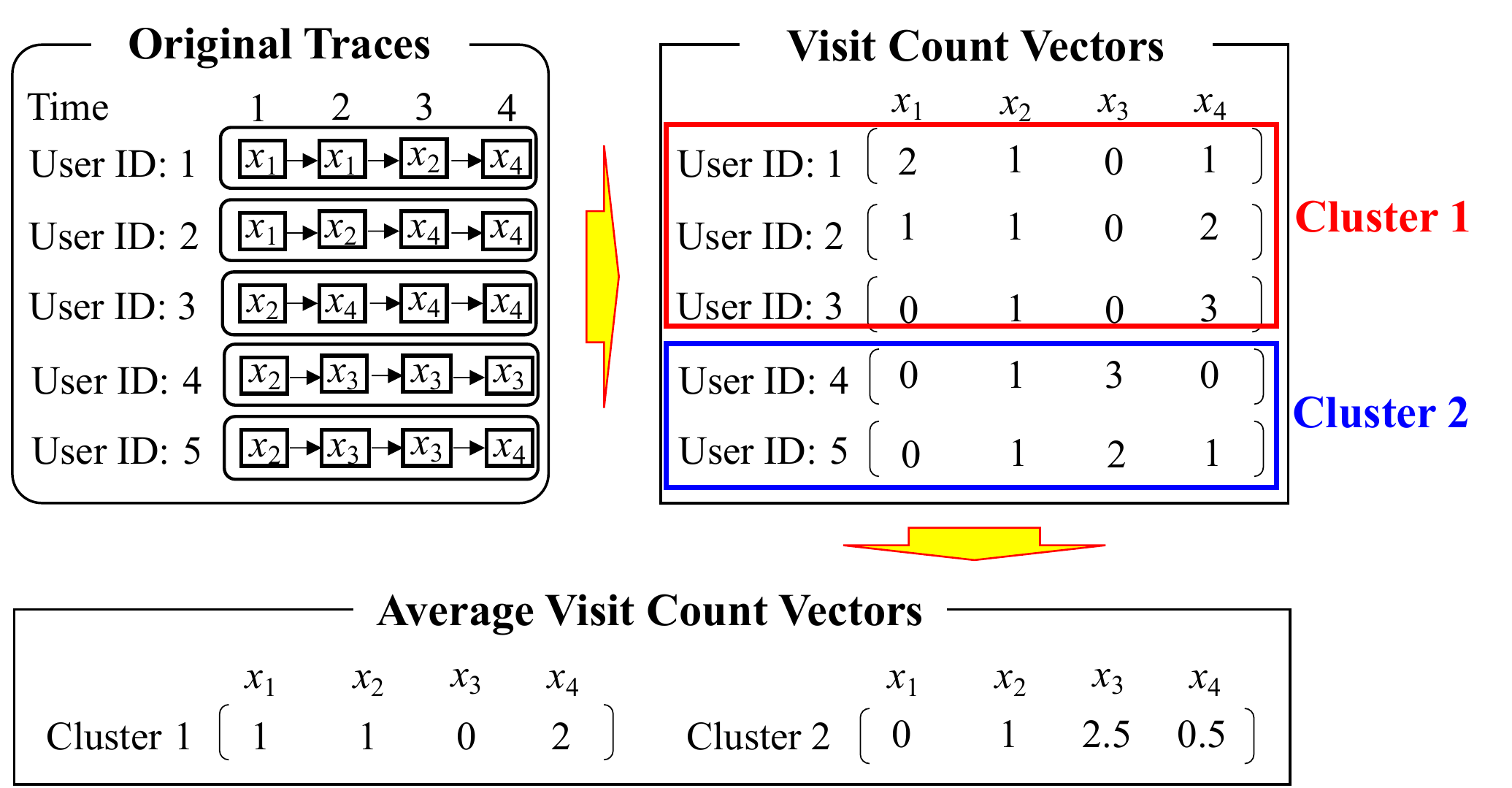}
\vspace{-3mm}
\caption{Clusters of visit count vectors.} 
\label{fig:cluster_visit_count}
\end{figure}

Specifically, in step (i), it clusters users based on visit count vectors using the $k$-means clustering algorithm, where the number $k$ of clusters is $k=100$. 
In step (ii), it calculates the average visit count vector within each cluster and adds noise to regions of each trace to move its visit count vector close to the average visit count vector. 
Steps (ii) and (iii) are performed under the constraint of the utility requirement (utility score $\geq 0.7$). 
Figure~\ref{fig:cluster_visit_count} shows a simple example of the clusters ($k=2$) and the average visit count vectors. 

\smallskip
\noindent{\textbf{\colorB{Difference from Existing Algorithms.}}}~~\colorB{The best algorithm is based on $k$-means clustering and is somewhat similar to $k$-anonymity based trace obfuscation \cite{Bettini_Springer09,Chow_SIGKDD11,Gedik_TMC08}. 
However, it differs from \cite{Bettini_Springer09,Chow_SIGKDD11,Gedik_TMC08} in that the best algorithm does \textit{not} provide $k$-anonymity that requires too much noise for long traces. 
\colorR{It is well known that both $k$-anonymity and DP could destroy utility for long traces \cite{Andres_CCS13,Bettini_Springer09,Pyrgelis_PoPETs17}.} 
The best algorithm avoids this issue by obfuscating traces within each cluster so that they are similar rather than identical. 
The fact that this algorithm won first place suggests that we need to look beyond popular notions such as $k$-anonymity and DP to achieve high utility for long traces\footnote{\colorR{Because the best algorithm uses $k$-means clustering, it might provide some theoretical guarantees, such as a relaxation of $k$-anonymity.} }.}

\subsection{Best Attack Algorithms}
\label{sub:ri_ti_teams}
We also explain attack algorithms for re-identification and trace inference developed by a team that won first place in re-identification (i.e., the best re-identification award) and third place in trace inference. 
The source code is also published in \cite{PWSCup2019}\footnotemark[6]. 
Although a team that won first place in trace inference 
is different, we omit its 
algorithm because the sum of privacy scores against the other teams is similar between the first to fourth teams. 

The best attack team is from a company. 
The age range is 40s. 
The team members have also participated in past PWS Cups. 

\smallskip
\noindent{\textbf{Re-identification Algorithm.}}~~\colorB{The best attack algorithm introduces a \textit{fuzzy counting} technique as a basic strategy. 
The fuzzy counting technique counts each region in the reference trace and its surrounding regions ($8$ regions) to construct an attack model.}

\begin{figure}
\centering
\includegraphics[width=0.9\linewidth]{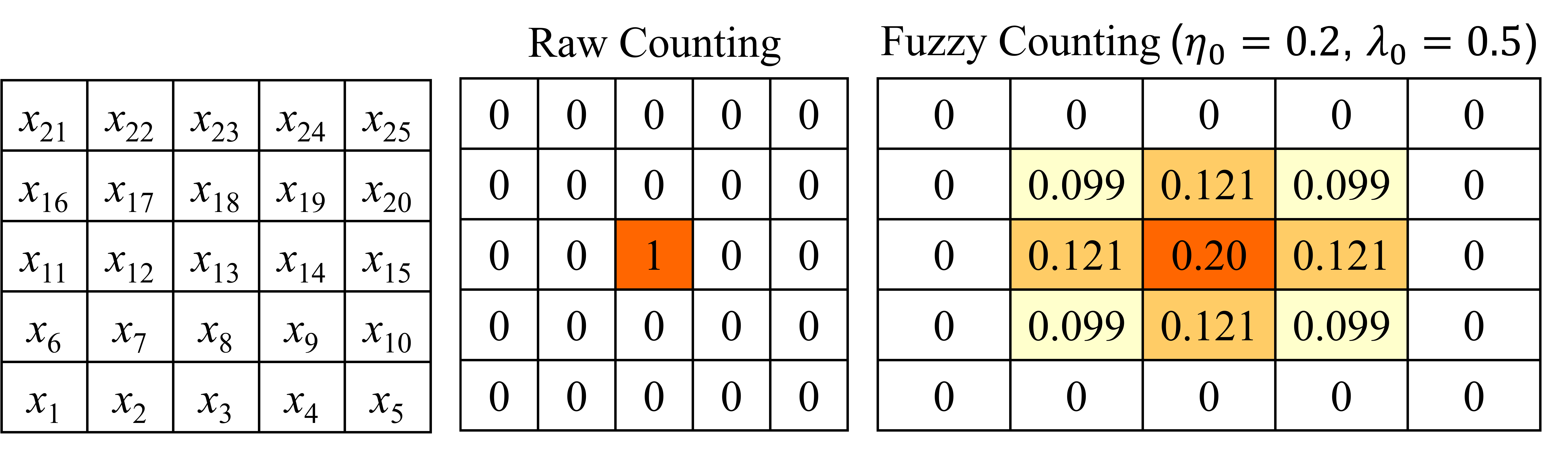}
\conference{\vspace{-3mm}}\arxiv{\vspace{-3mm}}
\caption{Fuzzy counting. 
In this example, the target region is $x_{13}$. 
The raw counting technique simply counts the target region. 
The fuzzy counting technique increases 
counts for surrounding regions, e.g.,  
count for $x_8$ by $0.121$. 
} 
\label{fig:fuzzy_count}
\end{figure}

Specifically, when generating a visit count vector for each user from her reference trace, this technique counts each region in the reference trace (referred to as a \textit{target region}) and its surrounding regions fuzzily. 
The fuzzy count for each region is determined by an exponential decay function $h(d) = \eta_o e^{-\lambda_0 d}$, where $\eta_0$ and $\lambda_0$ are constants and $d$ is the Euclidean distance 
from the target region. 
The Euclidean distance is normalized so that the distance between two neighbor regions is $1$. 
Figure~\ref{fig:fuzzy_count} shows an example of the fuzzy counting when $\eta_0=0.2$ and $\lambda_0=0.5$. 

From each visit count vector, it generates a \textit{term frequency–inverse document frequency (TF-IDF)} style feature vector to weigh unpopular regions more than popular regions. 
Specifically, 
let $\gamma_{i,j} \in \nngreals$ be a count of region $x_i$ in user $u_j$'s trace, and $\xi_i \in \nnints$ be the number of users whose trace includes region $x_i$ ($\xi_i \leq m = 2000$). 
Then it calculates the $i$-th element of the 1024-dim feature vector of user $u_j$ by TF $\cdot$ IDF, where (TF, IDF) 
$= (\gamma_{i,j}, \log \frac{m}{\xi_i})$, 
$(\gamma_{i,j}, 1)$, 
$(\log (1+\frac{m}{\xi_i}), \log \frac{m}{\xi_i})$, or 
$(\log (1+\frac{m}{\xi_i}), 1)$. 
Based on the feature vectors, it finds a user ID for each pseudonym via the 1-nearest neighbor search.
Optimal values of $\eta_0$, $\lambda_0$, and (TF, IDF) were determined 
using 
sample traces 
described in Section~\ref{sub:details_contest}. 
The optimal values were as follows: $\eta_0=0.33$, $\lambda_0=1$, and (TF, IDF) $= (\log (1+\frac{m}{\xi_i}), 1)$.

\smallskip
\noindent{\textbf{Trace Inference Algorithm.}}~~The trace inference algorithm of this team first re-identifies traces using the re-identification algorithm explained above. 
Then it de-obfuscates the original regions for each re-identified trace 
in a similar way to \textano{VisitProb-T} with 
an additional technique -- \textit{replacing frequent regions}.
\colorB{The basic idea of this technique is that if a user frequently visits a region in reference traces, then she also frequently visits the region in original traces.} 

Specifically, 
from reference traces, this technique calculates 
a region with the largest visit-count 
for each user and each time from 8:00 to 18:00. 
Because the reference trace length is $20$ days, there are $20$ visit-counts in total for each user and each time. 
If the 
visit-count 
exceeds a threshold (determined using sample traces), it regards the region as \textit{frequent}. 
Finally, it replaces a region with the frequent region (if any) for each user and each time in the inferred traces.

\smallskip
\noindent{\textbf{\colorB{Difference from Existing Algorithms.}}}~~The best attack algorithm uses a fuzzy counting technique as a basic strategy. Existing work \cite{Gambs_JCSS14,Mulder_WPES08,Shokri_SP11} and our sample algorithms in 
Section~\ref{sub:pre_setup} 
only count each region in the reference trace to construct an attack model. 
Thus, the fuzzy counting technique is more robust to small changes in the locations. 
In fact, the best algorithm provides much better attack accuracy than our sample algorithms. 

Fuzzy counting is also simple and much more efficient than complicated attacks such as \cite{Murakami_PoPETs17,Murakami_TIFS17}. 
Specifically, 
let $\eta \in \nats$ be the total number of regions. 
Then, the time complexity of fuzzy counting is 
$O(m (l_r + \eta))$, 
whereas that of \cite{Murakami_PoPETs17,Murakami_TIFS17} is $O(m l_r \eta^2)$. 

\subsection{\colorB{Relationship with \colorR{the Expected Error}}}
\label{sub:other_metrics}
\colorB{In our contest, we used a trace inference privacy score $s_T\ttp$ with hospital weight $=10$. 
Below, we analyze the relationship between the expected error \cite{Shokri_SP11} and our privacy scores $s_T\ttp$ with various hospital weights.} 

\begin{figure}
\centering
\includegraphics[width=0.99\linewidth]{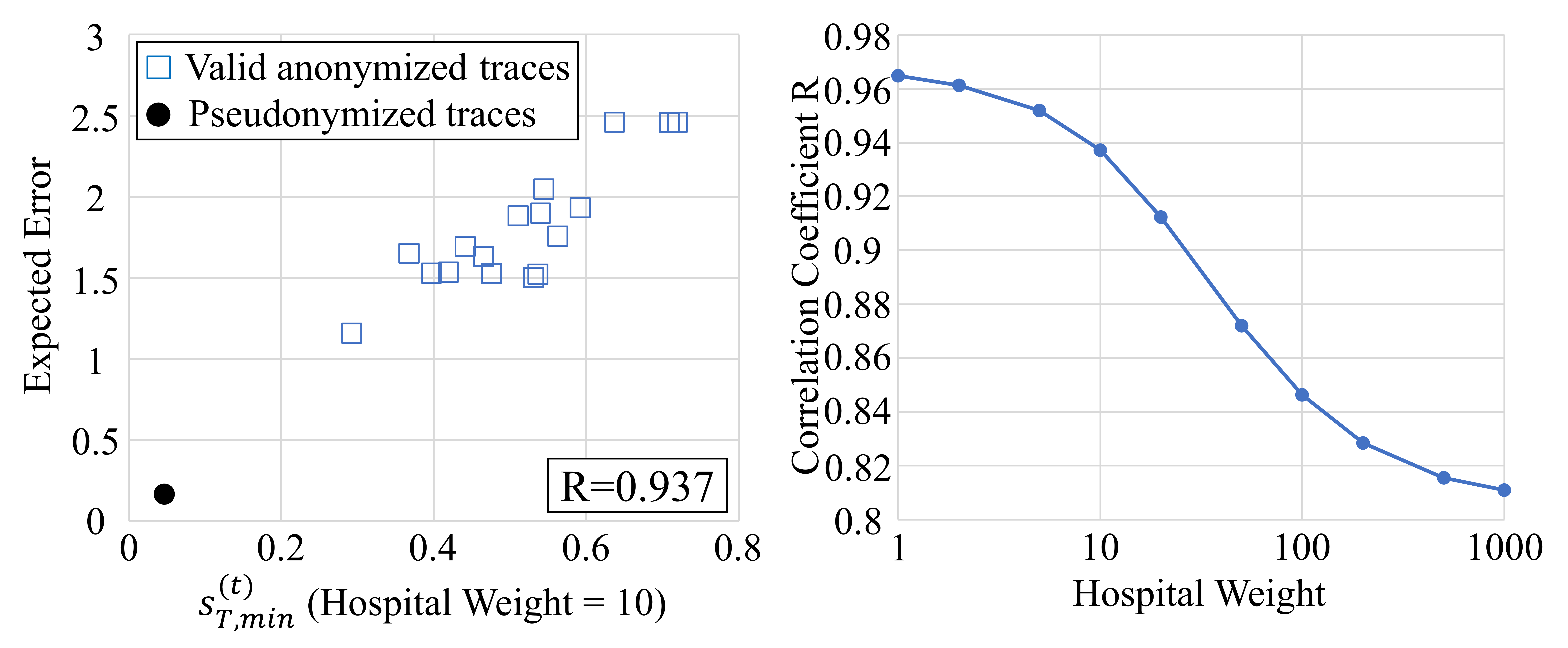}
\vspace{-4mm}
\caption{\colorB{Relationship between 
\colorR{our trace inference privacy scores with various hospital weights and 
the expected error \cite{Shokri_SP11}}. 
R is the correlation coefficient.}
} 
\label{fig:res_other_metrics}
\end{figure}

\colorB{The left panel of Figure~\ref{fig:res_other_metrics} shows the relationship between 
\colorR{our privacy score with hospital weight $=10$ and 
the expected error}. 
Here, we used the valid anonymized traces of the $17$ teams and the pseudonymized traces in our contest. 
We observe that 
our privacy score 
with hospital weight $=10$ is highly correlated with the expected error -- 
the correlation coefficient is R $=0.937$.} 

\colorR{The right panel of Figure~\ref{fig:res_other_metrics} shows the relationship between the correlation coefficient R and the hospital weight. 
We observe that as the hospital weight increases from $10$, R rapidly decreases; e.g., R $=0.846$ and $0.811$ when hospital weight $=100$ and $1000$, respectively. 
This means that if we choose such large weights, the best anonymization algorithm loses its versatility -- it may not be useful when the expected error is used as a privacy metric. 
In contrast, the best anonymization algorithm in our contest carefully handles sensitive regions (as hospital weight $\gg 1$) and also provides the largest expected error, as shown in the left panel of Figure~\ref{fig:res_other_metrics}.}

\subsection{Utility in Our Contest}
\label{sub:utility_contest}
\colorB{We finally analyzed the utility of the valid anonymized traces of the $17$ teams for various applications as follows.} 

\smallskip
\noindent{\textbf{\colorB{POI Recommendation.}}}~~\colorB{As described in Section~\ref{sub:details_contest}, anonymized traces are useful for POI recommendation \cite{Cheng_IJCAI13,Feng_IJCAI15,Liu_VLDB17} in the intermediate model. 
In our analysis, we considered the following successive personalized POI recommendation.} 
Suppose that a user is interested in POIs within a radius of $r_1$ km from each location in her original trace (referred to as \textit{nearby POIs}). 
To recommend the nearby POIs, the LBS provider sends all POIs within a radius of $r_2$ km from each location in the anonymized trace to the user through the intermediate server. 
Then the client application makes a recommendation of POIs based on the received POIs. 

Note that the client application knows the original locations. 
Therefore, it can filter the received POIs, i.e., exclude POIs outside of the radius of $r_1$ from the original locations. 
Thus, $r_2$ can be set to be larger than $r_1$ to increase the accuracy at the expense of higher communication cost \cite{Andres_CCS13}.

We extracted POIs in the ``food'' category from the SNS-based people flow data \cite{SNS_people_flow} ($4692$ POIs in total). 
We set $r_1 = 1$ km and $r_2 = 2$ km. 
Then we evaluated the proportion of 
nearby POIs included in the received POIs to the total number of nearby POIs and averaged it over all locations in the original traces (denoted by \textutl{POI Accuracy}). 

\smallskip
\noindent{\textbf{\colorB{Geo-Data Analysis.}}}~~\colorB{We also evaluated the utility for geo-data analysis, such as mining popular POIs \cite{Zheng_WWW09} and modeling human mobility patterns \cite{Liu_CIKM13,Song_TMC06}. To this end, we evaluated a population distribution and a transition matrix in the same way as \cite{Bindschaedler_SP16,Murakami_PoPETs21}.} 

\colorB{The population distribution is a basic statistical feature for mining popular POIs \cite{Zheng_WWW09}.} 
For each time from 8:00 to 18:00, we calculated a frequency distribution ($1024$-dim vector) of the original traces and that of the anonymized traces. 
For each time, we extracted the top $50$ POIs whose frequencies in the original traces were the largest and regarded the frequencies of the remaining POIs as $0$. 
\colorB{Here, we followed \cite{Bindschaedler_SP16} and selected the top $50$ locations.} 
Then, we evaluated the average total variance between the two time-dependent population 
distributions over all time (\textutl{TP-TV-Top50}). 

The transition matrix is a basic feature for modeling human mobility patterns \cite{Liu_CIKM13,Song_TMC06}. 
We calculated an average transition matrix ($1024 \times 1024$ matrix) over all users and all time. 
We calculated the transition matrix of the original traces and that of the anonymized traces. 
Each row of the transition matrix represents a conditional distribution. 
Thus, \colorB{we evaluated the Earth Mover's Distance (EMD) between the two conditional distributions in the same way as \cite{Bindschaedler_SP16}} and took an average over all rows (\textutl{TM-EMD}). 

Since the two-dimensional EMD is computationally expensive, we calculated the sliced 1-Wasserstein distance \cite{Bonneel_JMIV15,Kolouri_CVPR18}. 
The sliced 1-Wasserstein distance generates a lot of random projections of the 2D distributions to 1D distributions. 
Then it calculates the average EMD between the 1D distributions.

\begin{figure}
\centering
\includegraphics[width=0.99\linewidth]{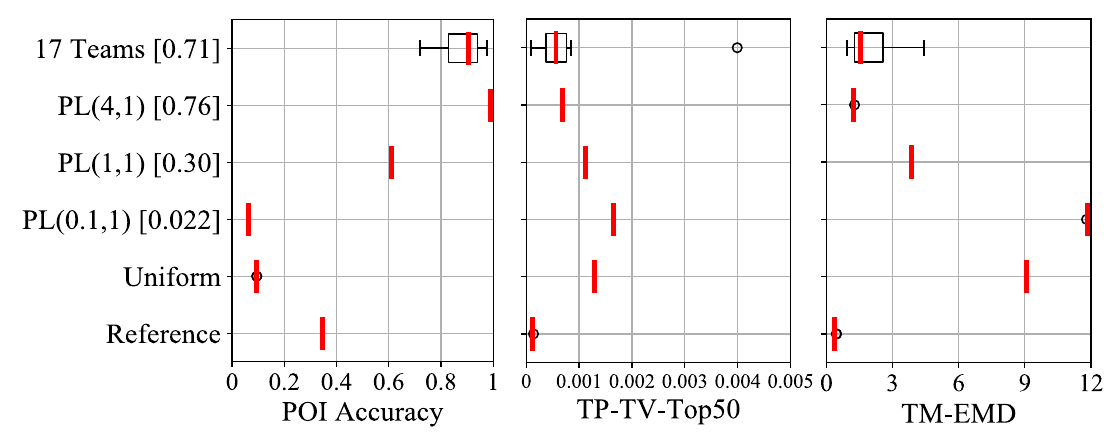}
\conference{\vspace{-3mm}}\arxiv{\vspace{-3mm}}
\caption{Box plots of $17$ points (utility values of the $17$ teams). 
The number in the square bracket represents the average utility score $s_U\tth$. 
The red line represents the median. 
The variance is very small for 
\textano{PL}, \textano{Uniform}, and \textano{Reference} because the same obfuscation method is used for all of the original traces. 
Higher is better for \textutl{POI Accuracy}. 
Lower is better for the others.} 
\label{fig:util_res}
\end{figure}

\smallskip
\noindent{\textbf{\colorB{Results.}}}~~Figure~\ref{fig:util_res} shows the box plots of $17$ points (utility values of the $17$ teams) for each utility metric, 
where 
``\textano{17 Teams}'' represents the valid anonymized traces submitted by the $17$ teams. 
We also evaluated \textano{PL($4, 1$)}, \textano{PL($1, 1$)}, and  \textano{PL($0.1, 1$)} applied to the original traces of the $17$ teams. 
\textano{Uniform} is the utility when all locations in the anonymized traces are independently sampled from a uniform distribution. 
\textano{Reference} is the utility when the reference traces are used as anonymized traces. 

Figure~\ref{fig:util_res} shows that \textano{17 Teams} and \textano{PL($4, 1$)} 
(both of which satisfy $s_U\tth > 0.7$) provide very high utility for \textutl{POI Accuracy} (the median is $0.9$ or more). 
This is because the POI recommendation task explained above requires each location in the anonymized trace to be close to the corresponding location in the original trace. 
Since the utility score in our contest is based on this requirement, it is inherently suitable for POI recommendation based on anonymized traces. 
\colorB{Figure~\ref{fig:util_res} also shows that the POI accuracy of \textano{Reference} is low. 
This means that preserving only statistical information of the original traces is not sufficient for POI recommendation.} 

Figure~\ref{fig:util_res} also shows that \textano{17 Teams} and \textano{PL($4, 1$)} provide almost the same performance as \textano{Reference} for \textutl{TP-TV-Top50} and \textutl{TM-EMD}, which means that statistical information is also well preserved when 
$s_U\tth > 0.7$. 
Therefore, our utility score can be used as a simple guideline to achieve high utility in geo-data analysis. 

\colorB{In summary, 
the valid anonymized traces 
are useful for various applications, including POI recommendation and geo-data analysis (e.g., mining popular POIs, modeling human mobility patterns).}

\section{Conclusion}
\label{sec:conclusion}
We designed and held 
a location trace anonymization contest 
that deals with a long trace and fine-grained locations. 
We 
showed through the contest 
that an anonymization method secure against trace inference is also secure against re-identification 
in a situation where both defense and attack compete together. 
\colorB{We also showed that the anonymized traces in our contest are useful for various applications including POI recommendation and geo-data analysis.}

\section*{Acknowledgments}
The authors would like to thank S\'{e}bastien Gambs (UQAM) for technical comments on this paper. 
The authors would like to thank Takuma Nakagawa (NSSOL) for providing the information on the anonymization algorithm in 
Section~\ref{sub:best_algorithm} and 
all teams for participating in our contest. 
The authors would also like to thank anonymous reviewers for helpful suggestions. 
This study was supported in part by JSPS KAKENHI JP18H04099 and JP19H04113. 

\bibliographystyle{ACM-Reference-Format}
\bibliography{main_conf}

\appendix
\section{Location Synthesizer for a Contest}
\label{sec:synthesizer}
\colorB{Below, we explain the details of our 
location synthesizer. 
Our location synthesizer extends the location synthesizer in \cite{Murakami_PoPETs21} called privacy-preserving multiple tensor factorization (\textsyn{PPMTF}) to 
have diversity. 
Therefore, we denote our location synthesizer by \textsyn{PPMTF+}. 
We first briefly review \textsyn{PPMTF} (the part most relevant to our location synthesizer). 
Then, we explain how our location synthesizer \textsyn{PPMTF+} extends \textsyn{PPMTF} to have diversity.} 

\smallskip
\noindent{\textbf{\colorB{PPMTF.}}}~~\colorB{\textsyn{PPMTF} trains a feature vector of each training user 
from training traces and generates a synthetic trace from the feature vector. 
Let $n \in \nats$ be the number of training users and $z \in \nats$ be the dimension of the feature vector. 
Let $\bmA \in \reals^{n \times z}$ be the feature vectors of all training users, and $\bma_i \in \reals^z$ be the $i$-th row of $\bmA$. 
$\bma_i$ represents a feature vector of the $i$-th training user. 
Each column in $\bmA$ represents a \textit{cluster} of users, such as ``those who go to a bar at night'' and ``those who go to a park at noon.'' 
For example, assume that the first column of $\bmA$ represents a cluster of users who go to a bar at night. 
Then, a user who has a large value in the first element of her feature vector $\bma_i$ tends to go to bar at night in her trace. 
\textsyn{PPMTF} automatically finds such user clusters ($z$ clusters in total) from training traces.} 

\colorB{More specifically, \textsyn{PPMTF} trains feature vectors and generates traces as follows. 
First, \textsyn{PPMTF} assumes that each feature vector is independently generated from a multivariate normal distribution: 
\begin{align}
p(\bmA | \PsiA) &=  \textstyle\prod_{i=1}^{n} \calN(\bma_i | \muA, \LamA\inv), 
\label{eq:bmA}
\end{align}
where 
$\muA \in \reals^z$ is a mean vector and $\LamA \in \reals^{z \times z}$ is a precision (inverse covariance) matrix. 
$\calN(\bma_i | \muA, \LamA\inv)$ denotes the probability of $\bma_i$ in the normal distribution with mean vector $\muA$ and covariance matrix $\LamA\inv$. 
Let $\PsiA = (\muA, \LamA)$. 
$\PsiA$ forms a feature vector distribution. 
\textsyn{PPMTF} trains $\PsiA$ from training traces via posterior sampling \cite{Wang_ICML15}, which samples a parameter from its posterior distribution given training data. 
Then, \textsyn{PPMTF} trains $\bmA$ based on $\PsiA$ and training traces. 
Finally, \textsyn{PPMTF} generates a synthetic trace that resembles the $i$-th training user's trace based on $\bma_i$.} 

\colorB{\textsyn{PPMTF} provides high utility in that it preserves various statistical features of training traces such as a distribution of visit-fractions \cite{Do_TMC13,Ye_KDD11}, time-dependent population distribution \cite{Zheng_WWW09}, and transition matrix \cite{Liu_CIKM13,Song_TMC06}. 
\textsyn{PPMTF} also protects privacy of training users. 
Specifically, the authors in \cite{Murakami_PoPETs21} show through experiments that the synthetic traces are secure against re-identification attacks \cite{Gambs_JCSS14,Mulder_WPES08,Murakami_PoPETs17,Shokri_SP11} and membership inference attacks \cite{Jayaraman_USENIX19,Shokri_SP17}. 
See \cite{Murakami_PoPETs21} for more details of \textsyn{PPMTF}.} 

\smallskip
\noindent{\textbf{\colorB{Our Location Synthesizer \textsyn{PPMTF+}.}}}~~\colorB{After training $\PsiA$ and $\bmA$ from training traces, \textsyn{PPMTF} generates traces from $\bmA$. 
In contrast, our location synthesizer \textsyn{PPMTF+} discards $\bmA$ and randomly samples \textit{new} feature vectors from $\PsiA$.} 

\begin{figure}[t]
\centering
\includegraphics[width=0.99\linewidth]{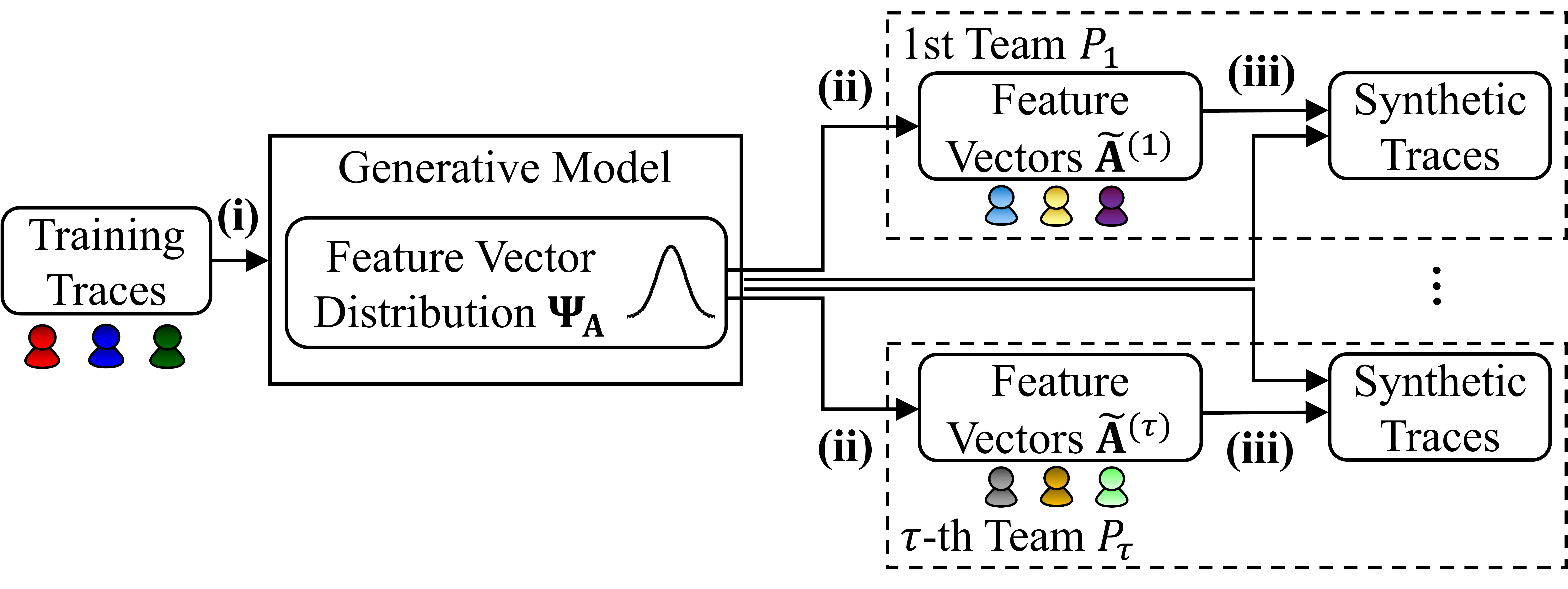}
\conference{\vspace{-3mm}}\arxiv{\vspace{-3mm}}
\caption{Our location synthesizer \textsyn{PPMTF+}. It consists of the following three steps: (i) training a generative model, (ii) sampling new feature vectors, and (iii) generating synthetic traces. 
}
\label{fig:synthesizer_generate}
\end{figure}

Figure~\ref{fig:synthesizer_generate} shows our synthesizer \textsyn{PPMTF+}. 
The number $m$ of virtual users can be different from the number $n$ of training users. 
We first train a generative model in the same way as \textsyn{PPMTF}. 
Then, we randomly sample a \textit{new} matrix $\tbmA^{(t)} \in \reals^{m \times z}$ ($m$ feature vectors) for team $P_t$ from $\PsiA$. 
Specifically, let $\tbma_i\tth \in \reals^z$ be the $i$-th row of $\tbmA\tth$, i.e., the $i$-th virtual user's feature vector in team $P_t$. 
We randomly sample $\tbma_i\tth$ as follows: 
\begin{align}
\tbma_i\tth \sim \calN( \cdot | \muA, \LamA\inv), 
\label{eq:tbma_i}
\end{align}
where $\calN( \cdot | \muA, \LamA\inv)$ denotes the normal distribution with mean vector $\muA$ and covariance matrix $\LamA\inv$. 
Note that we do not use training traces 
in (\ref{eq:tbma_i}). 
\textsyn{PPMTF} trains $\bmA$ based on $\PsiA$ and training traces so that $\bmA$ preserves the feature of training users. 
In contrast, we randomly sample $\tbma_i\tth$ based on $\PsiA$, independently of training traces. 
\colorB{After sampling $\tbmA^{(t)}$, we generate synthetic traces by replacing $\bmA$ with $\tbmA^{(t)}$ in \textsyn{PPMTF} (see \cite{Murakami_PoPETs21} for how to generate synthetic traces from the generative model in \textsyn{PPMTF}).} 

Since each feature vector $\tbma_i\tth$ is independently and randomly sampled, each virtual user has a different feature vector than the training users and the other virtual users. 
This yields the \textit{diversity}. 
In addition, both $\bma_i$ and $\tbma_i\tth$ follow the normal distribution with parameter $\PsiA = (\muA, \LamA)$ (see (\ref{eq:bmA}) and (\ref{eq:tbma_i})). 
Thus, $\tbmA\tth$ preserves very similar statistical features to $\bmA$. 
Consequently, our synthetic traces provide high utility in terms of various statistical features in the same way as \textsyn{PPMTF}. 

Our synthesizer \textsyn{PPMTF+} is also as scalable as \textsyn{PPMTF} because sampling $\tbmA^{(t)}$ takes little time, e.g., much less than one second even using a laptop. 
For example, \textsyn{PPMTF+} can synthesize traces of about $200000$ users within one day in the same way as \textsyn{PPMTF} \cite{Murakami_PoPETs21}.

\smallskip
\noindent{\textbf{Remark.}}~~In our contest, we kept the algorithm of our synthesizer secret from all teams. 
However, we argue that a contest would be interesting even if the algorithm is made public. 
Specifically, if each team $P_t$ keeps feature vectors $\tbmA^{(t)}$  private, the other teams cannot obtain information on how each virtual user in team $P_t$ behaves; e.g., some may often go to restaurants, and others may commute by train. 
Thus, both attacks and defenses are still very challenging and interesting.

\section{Diversity and Utility of Our Synthesizers}
\label{sec:exp_syn}

\subsection{Experimental Set-up}
\label{sub:exp_setup}

\noindent{\textbf{Dataset.}}~~We evaluated the diversity and utility of 
\textsyn{PPMTF+} in Appendix~\ref{sec:synthesizer} 
using 
the Foursquare dataset 
in 
\cite{Yang_WWW19}. 
Following \cite{Yang_WWW19}, we selected six cities with many check-ins and with cultural diversity: 
Istanbul (\textdat{IST}), 
Jakarta (\textdat{JK}), 
New York City (\textdat{NYC}), 
Kuala Lumpur (\textdat{KL}), 
San Paulo (\textdat{SP}), and 
Tokyo (\textdat{TKY}). 

For each city, we selected $1000$ popular POIs for which the number of visits from all users was the largest. 
Thus, the number of locations is $1000$ in our experiments. 
We set the time interval between two temporally-continuous events 
to one hour by rounding down minutes. 
Then we set a time slot \cite{Murakami_PoPETs21}, a time resolution at which we want to preserve a population distribution of training traces, to two hours, i.e., $12$ time slots in total. 
For each city, we randomly selected $80\%$ of users as training users and used the remaining users 
as testing users. 
The traces of testing users were used for evaluating a baseline (the utility of real traces). 
The numbers of training users in \textdat{IST}, \textdat{JK}, \textdat{NYC}, \textdat{KL}, \textdat{SP}, and 
\textdat{TKY} were 
$n = 219793$, $83325$, $52432$, $51189$, $42100$, and $32056$, respectively. 

\smallskip
\noindent{\textbf{Synthesizers.}}~~We generated synthetic traces using 
\textsyn{PPMTF+}. 
We set 
the dimension $z$ of the feature vector to $z=16$ 
(in the same way as \cite{Murakami_PoPETs21}) 
and 
the number $\tau$ of teams to $\tau=2$. 
Then we generated 
the same number of 
virtual users' feature vectors 
as training users 
($m = n$). 
For each virtual user, we generated 
one synthetic trace with the length of 
$20$ days. 

For comparison, we 
generated 
the same amount of synthetic traces using \textsyn{PPMTF} \cite{Murakami_PoPETs21}. 
\colorB{As another baseline, we also evaluated 
the synthetic data generator (denoted by \textsyn{SGD}) in \cite{Bindschaedler_VLDB17}. 
Specifically, we first calculated a transition matrix for each time slot 
($12 \times 1000 \times 1000$ elements in total) 
from training traces via maximum likelihood estimation. 
Then we generated synthetic traces by randomly sampling locations using the transition matrix. 
This method is a special case of the synthetic data generator in \cite{Bindschaedler_VLDB17} where the generative model is independent of the input data record (see \cite{Murakami_PoPETs21} for details).} 

\textsyn{SGD} generates traces only based on parameters common to all users. 
\colorB{Therefore, it does not preserve user-specific features, as shown in our experiments. 
Note that the user-specific features are necessary for an anonymization contest, because otherwise the adversary cannot re-identify traces. 
In other words, the adversary needs some user-specific features as background knowledge to re-identify traces.} 
We do not evaluate other 
synthesizers such as \cite{Chen_CCS12,Chen_KDD12,He_VLDB15}, 
because they also generate traces only based on parameters common to all users and have the same issue. 

We also note that a location synthesizer in \cite{Bindschaedler_SP16} lacks scalability and cannot be applied to the Foursquare dataset; e.g., it requires over four years to generate traces in \textdat{IST} even using 
a supercomputer 
\cite{Murakami_PoPETs21}. 

\smallskip
\noindent{\textbf{Diversity Metrics.}}~~For diversity, 
we evaluated whether the $i$-th synthetic trace for the 1st team is similar to the $i$-th synthetic trace for the 2nd team. 

Specifically, we first randomly selected 
$2000$ virtual users from $m$ virtual users. 
Without loss of generality, we denote the selected user IDs by $1, \cdots, 2000$. 
We calculated a population distribution ($1000$-dim vector) for each 
virtual user. 
Then we evaluated the average total variance between the distribution of the 
$i$-th user in the first team and that of the 
$i$-th (resp.~$(1000+i)$-th) user in the second team ($1 \leq i \leq 1000$), which is denoted by \textutl{Same-TV} (resp.~\textutl{Diff-TV}). 
If \textutl{Same-TV} is smaller than \textutl{Diff-TV}, the synthesizer lacks diversity. 

Note that \textsyn{PPMTF+} clearly has diversity because it independently and randomly samples 
each 
feature vector 
by (\ref{eq:tbma_i}). 
The purpose here is to quantitatively show that \textsyn{PPMTF} lacks diversity. 

\smallskip
\noindent{\textbf{Utility Metrics.}}~~For utility, we evaluated 
a distribution of visit-fractions, 
time-dependent 
population distribution, and 
transition matrix in the same way as \cite{Murakami_PoPETs21}. 

\begin{figure}[t]
\centering
\includegraphics[width=0.9\linewidth]{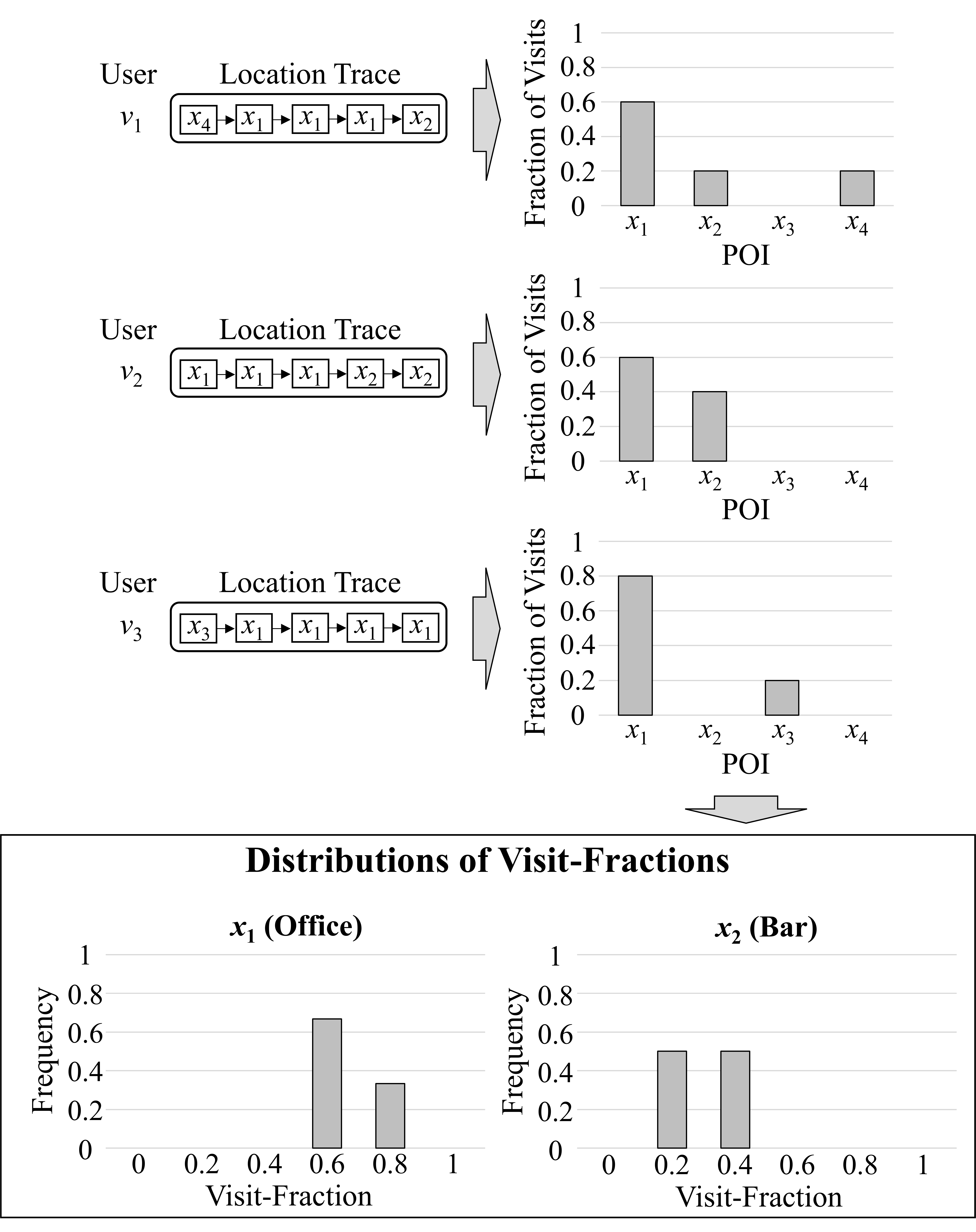}
\conference{\vspace{-2mm}}\arxiv{\vspace{-2mm}}
\caption{\colorB{Example of calculating a distribution of visit-fractions from location traces.}}
\label{fig:visit-fraction}
\end{figure}

The distribution of visit-fractions is a key feature for auto-tagging POI categories (e.g., restaurants, hotels)~\cite{Do_TMC13,Ye_KDD11}. 
For example, many people spend $60\%$ of the time at home and $20\%$ of the time at work/school \cite{Do_TMC13}. 
To evaluate such a feature, we did the following. 
For each training user, 
we computed a fraction of visits for each POI. 
Then we computed a distribution of visit-fractions for each POI by dividing the fraction into $24$ bins: $(0,\frac{1}{24}], (\frac{1}{24},\frac{2}{24}],\cdots,(\frac{23}{24},1)$. 
\colorB{Figure~\ref{fig:visit-fraction} shows an example of 
calculating a distribution of visit-fractions for each of POIs $x_1$ (office) and $x_2$ (bar). 
In this example, the fraction is divided into bins with a length of $0.2$, and people tend to stay in $x_1$ (office) longer than $x_2$ (bar).} 

We computed 
a distribution of visit-fractions 
for each POI 
from synthetic traces in the same way. 
Finally, we evaluated the total variance between the two distributions for each POI and took the average over all POIs (denoted by \textutl{VF-TV}).

\colorB{For the population distribution and the transition matrix, we evaluated \textutl{TP-TV-Top50} and \textutl{TM-EMD} in Section~\ref{sub:utility_contest}, respectively.}

\subsection{Experimental Results}
\label{sub:exp_res}

\begin{figure}[t]
\centering
\includegraphics[width=0.99\linewidth]{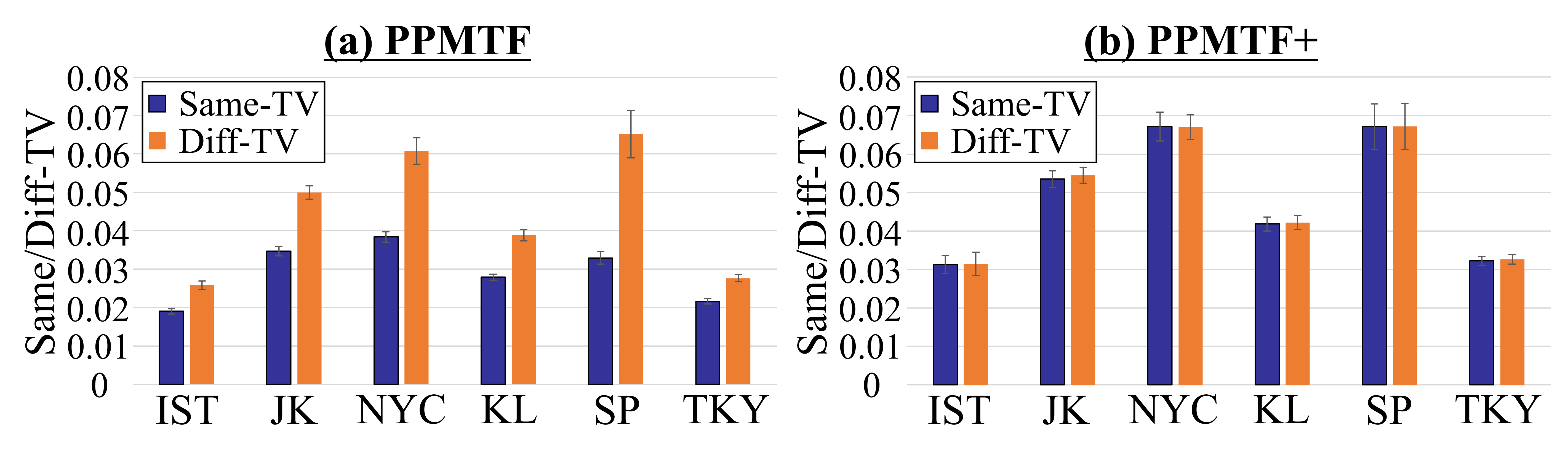}
\conference{\vspace{-3mm}}\arxiv{\vspace{-3mm}}
\caption{Diversity of \textsyn{PPMTF} and \textsyn{PPMTF+}. The error bar shows $0.1 \times$ the standard deviation 
over $1000$ pairs of users.}
\label{fig:res_syn_div}
\vspace{3mm}
\includegraphics[width=0.99\linewidth]{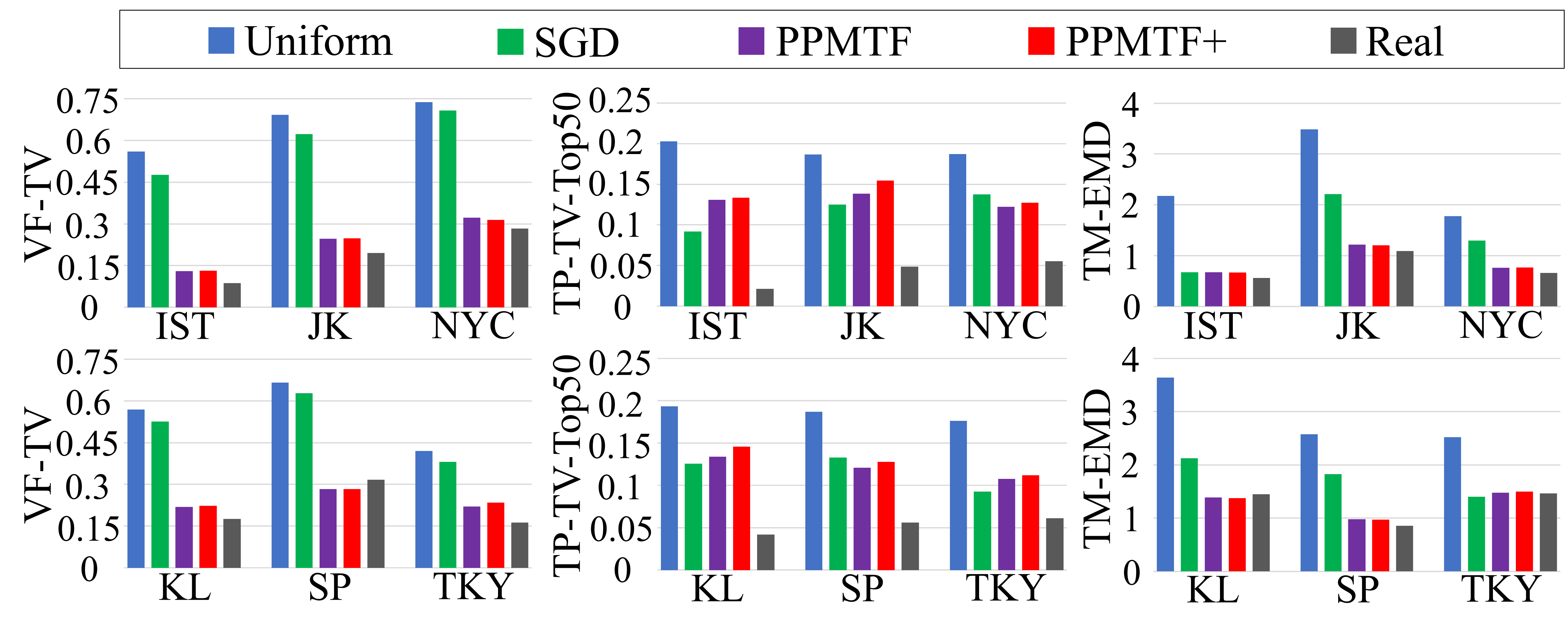}
\conference{\vspace{-3mm}}\arxiv{\vspace{-3mm}}
\caption{Utility of each location synthesizer. Lower is better in all metrics.}
\label{fig:res_syn_util}
\end{figure}

\noindent{\textbf{Diversity.}}~~We first evaluated the diversity of \textsyn{PPMTF+} and \textsyn{PPMTF}. 
Figure~\ref{fig:res_syn_div} shows the results. 
In \textsyn{PPMTF}, 
\textutl{Same-TV} is smaller than \textutl{Diff-TV}, which means that the $i$-th synthetic trace for the 1st team is similar to the $i$-th synthetic trace for the 2nd team.  
Thus, \textsyn{PPMTF} lacks diversity. 
In contrast, \textutl{Same-TV} is almost the same as \textutl{Diff-TV} (their difference is much smaller than $0.1$ times the standard deviation) in \textsyn{PPMTF+}. 
This is because \textsyn{PPMTF+} independently samples 
each 
feature vector. 

\smallskip
\noindent{\textbf{Utility.}}~~We next evaluated the utility. 
Figure~\ref{fig:res_syn_util} shows the results. 
Here, we evaluated each utility metric for each of $20$ days in the 1st team and averaged it over $20$ days. 
\textsyn{Uniform} is the utility when all locations in synthetic traces are independently sampled from a uniform distribution. 
\textsyn{Real} is the utility of the traces of testing users, i.e., real traces. 
Ideally, the utility of synthetic traces should be much better (smaller) than that of \textsyn{Uniform} and close to that of \textsyn{Real}. 

Figure~\ref{fig:res_syn_util} shows 
that \textsyn{SGD} provides poor utility (almost the same as \textsyn{Uniform}) in terms of \textutl{VF-TV}. 
This is because \textsyn{SGD} generates traces 
using 
parameters common to all users. 
Consequently, all users spend almost the same amount of time on each POI. 
In other words, \textsyn{SGD} cannot preserve 
user-specific features. 
\colorB{As explained above, the user-specific features are necessary for re-identification in an anonymization contest.} 

In contrast, 
\textsyn{PPMTF+} provides high utility in all utility metrics in the same way as \textsyn{PPMTF}. 
This is because 
feature vectors 
$\tbmA\tth$ in \textsyn{PPMTF+} preserve very similar statistical features to 
feature vectors 
$\bmA$ in \textsyn{PPMTF}. 

\smallskip
\noindent{\textbf{Summary.}}~~In summary, 
the existing synthesizers lack either diversity or \colorB{user-specific features}. 
Thus, they \textit{cannot} be applied to an anonymization contest in the partial-knowledge attacker model. 
In contrast, \textsyn{PPMTF+} provides both diversity and utility including user-specific features. 
\colorB{Thus, it fulfills the need for our contest.}

\section{Privacy of Our Synthesizer}
\label{app:privacy_location_synthsizer}

\noindent{\textbf{Experimental Set-up.}}~~To evaluate the privacy of 
\textsyn{PPMTF+} in Appendix~\ref{sec:synthesizer}, 
we used the SNS-based people flow data \cite{SNS_people_flow} (Tokyo) 
and processed the data 
in the same way as \cite{Murakami_PoPETs21}. 
There were $400$ regions and $n = 500$ training users (see \cite{Murakami_PoPETs21} for details). 
We generated synthetic traces using our location synthesizer \textsyn{PPMTF+} with $z=16$ and $\tau=1$. 
We generated synthetic traces of virtual users of the same number as the training users 
($m = n$). 
For each virtual user, we generated one synthetic trace with a length of ten days. 

For comparison, we also generated the same amount of synthetic traces using \textsyn{PPMTF}. 
Since we used a relatively small dataset ($n = 500$) in this experiment, we also evaluated the synthetic location traces generator in \cite{Bindschaedler_SP16} (denoted by \textsyn{SGLT}). 
We used the SGLT tool in \cite{Bindschaedler_SP16}. 
We set the number $c$ to semantic clusters to $c=50$ or $100$ because they provided the best performance. 
We set the other parameters in the same way as \cite{Murakami_PoPETs21}.

For utility, we evaluated \textutl{TP-TV-Top50} and \textutl{TM-EMD} in Section~\ref{sec:exp_syn}. 
We did not evaluate \textutl{VF-TV}, 
a utility measure for auto-tagging POI categories, 
because 
we used regions rather than POIs in this experiment.

For privacy, we considered two privacy attacks: 
re-identification (or de-anonymization) attack \cite{Gambs_JCSS14,Mulder_WPES08,Murakami_PoPETs17,Shokri_SP11} and membership inference attack \cite{Jayaraman_USENIX19,Shokri_SP17}. 
For the re-identification attack, we evaluated a 
re-identification rate. 
For the membership inference attack, we used the \textit{membership advantage} \cite{Jayaraman_USENIX19,Yeom_CSF18}, the difference between the true positive rate and the false positive rate, as a privacy metric. 
We used the implementation of these attacks in \cite{Murakami_PoPETs21} (see \cite{Murakami_PoPETs21} for details).

\begin{figure}[t]
\centering
\includegraphics[width=0.96\linewidth]{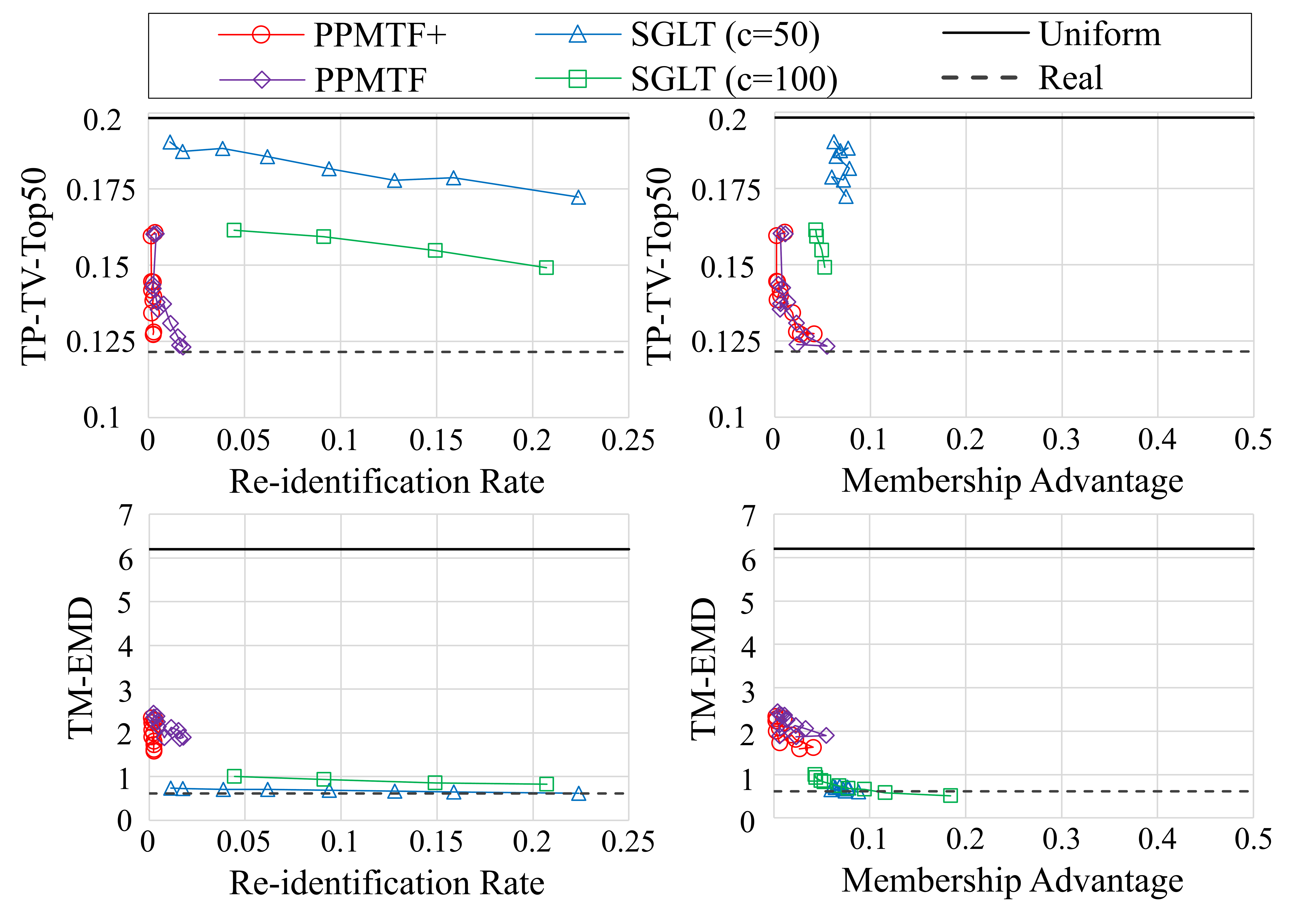}
\vspace{-3mm}
\caption{Privacy and utility of each location synthesizer. Lower is better in all metrics.}
\label{fig:res_syn_priv}
\end{figure}

\smallskip
\noindent{\textbf{Results.}}~~Figure~\ref{fig:res_syn_priv} shows the results. 
\textsyn{Uniform} represents the utility when all locations in synthetic traces are independently sampled from a uniform distribution.
\textsyn{Real} represents the utility of the testing traces. 

Figure~\ref{fig:res_syn_priv} shows that  \textsyn{PPMTF+} achieves 
almost the same re-identification rate as a random guess ($=0.002$). 
This is because \textsyn{PPMTF+} independently and randomly generates each virtual user from a distribution of user profiles. 
In other words, the virtual users are different from the training users, i.e., diversity. 
Figure~\ref{fig:res_syn_priv} also shows that \textsyn{PPMTF+} achieves 
almost the same membership advantage as \textsyn{PPMTF}. 

Note that 
\textutl{TM-EMD} of \textsyn{PPMTF+} is worse than that of \textutl{SGLT}. 
This is because \textsyn{PPMTF}, on which \textsyn{PPMTF+} is based, modifies the transition matrix of each training user 
via the Metropolis-Hastings algorithm (see \cite{Murakami_PoPETs21} for details). 
However, \textutl{TM-EMD} of \textsyn{PPMTF+} is still much lower than that of \textsyn{Uniform}, which means that \textsyn{PPMTF+} preserves the transition matrix well. 

In summary, our experimental results show that \textsyn{PPMTF+} achieves the same re-identification rate as a random guess and 
also has security against membership inference attacks in the same way as \textsyn{PPMTF}. 
Providing strong privacy guarantees 
such as DP \cite{Dwork_ICALP06,DP} is left for future work.

\section{Generation of Home Regions}
\label{app:home_regions}
\colorB{In our contest, we slightly modified \textsyn{PPMTF+} in Appendix~\ref{sec:synthesizer} 
so that 
each virtual user has her own home.} 

\colorB{We first extracted training traces from 6:00 to 18:00 for $10181$ users 
who have at least $10$ locations 
from the SNS-based people flow data. 
We trained our generative model from these training traces and sampled feature vectors of $m=2000$ virtual users in \textsyn{PPMTF+}. 
Then, we randomly selected a \textit{home region} for each virtual user from a population distribution at 6:00. 
\textsyn{PPMTF+} generates a synthetic trace of each user using a visit-count matrix, which includes a visit-count for each region and each time slot 
(see \cite{Murakami_PoPETs21} for details). 
Thus,  we increased visit-counts of the virtual user in her home region from 6:00 to 8:00 and 16:00 to 18:00 so that she stays at home in the morning and evening with high probability. 
After that, 
we generated a synthetic trace from 6:00 to 18:00 for 40 days for each virtual user using this modified synthesizer.} 

\begin{figure}[t]
\centering
\includegraphics[width=0.7\linewidth]{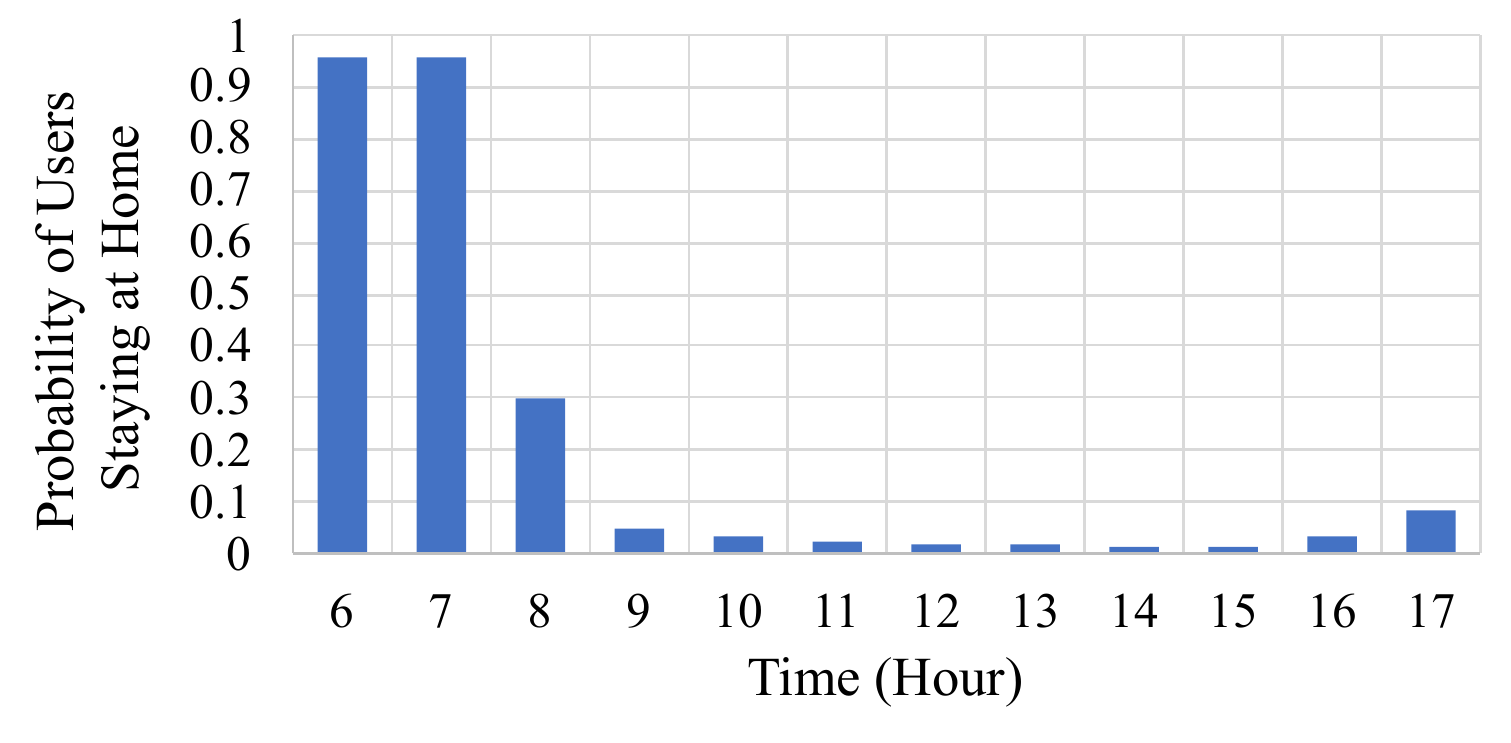}
\vspace{-5mm}
\caption{Probability of users staying at home.}
\label{fig:home}
\end{figure}

\colorB{Figure~\ref{fig:home} shows the probability of users staying at a home region. 
We also show in Figure~\ref{fig:regions} 
the population distribution of the 
synthetic traces in our contest at 12:00. 
These figures show 
that 
synthetic traces are generated so that 
each user stays at home in the morning and evening with high probability, while keeping the time-dependent population distribution 
of the training traces. 
We have also confirmed that our synthetic traces preserve the transition matrix of the training traces.} 

\colorB{After the generation of synthetic traces, we extracted locations from 8:00 to 18:00 in 30-minute intervals for each of 40 days and used them for our contest. 
Note that almost all virtual users stay at home from 6:00 to 8:00, as shown in Figure~\ref{fig:home}. 
Thus, if we include locations from 6:00 to 8:00 in reference traces, the adversary would know home regions of almost all virtual users. 
Because this is too informative for the adversary, we excluded locations from 6:00 to 8:00 in our contest.}

\section{Fairness in Our Contest}
\label{app:fairness_contest}

\colorB{We evaluated the fairness of our contest by calculating the variance of privacy scores as follows.} 
We anonymized the original traces of the $17$ teams using \textano{No Obfuscation}, \textano{MRLH($1,1,0.5$)}, \textano{RR($1$)}, or \textano{PL($1, 1$)}. 
Then we applied all the sample attack algorithms to the anonymized traces and evaluated the standard deviation (SD) of the privacy scores $s_{R,min}\tth$ and $s_{T,min}\tth$. 

Table~\ref{tab:SD_scores} shows the results. 
\colorB{The SD is very small when we apply \textano{RR($1$)}. 
This is because \textano{RR($1$)} rarely outputs the original region (with probability $0.0027$).} 
The SD is the largest when we do not apply any obfuscation algorithms (\textano{No Obfuscation}), in which case the SD is about $0.01$. 

\begin{table}[t]
\caption{Standard deviation (SD) of the minimum privacy scores 
$s_{R,min}\tth$ and $s_{T,min}\tth$ for the $17$ teams calculated by the sample anonymization and attack algorithms 
(NO: \textano{No Obfuscation}). } 
\conference{\vspace{-2mm}}
\label{tab:SD_scores}
\hbox to\hsize{\hfil
\arxiv{\footnotesize}
\begin{tabular}{c|c|c|c|c}
\hline 
	& \textano{NO} & \textano{MRLH($1,1,0.5$)} & \textano{RR($1$)} & \textano{PL($1, 1$)}\\
\hline 
    \hspace{-2mm} $s_{R,min}\tth$ \hspace{-2mm} & $0.0095$ & $0.0068$ & $0.00055$ & $0.0026$\\
\hline
    \hspace{-2mm} $s_{T,min}\tth$ \hspace{-2mm} & $0.011$ & $0.0033$ & $0.00015$ & $0.0010$\\
\hline
\end{tabular}
\hfil}
\end{table}

Table~\ref{tab:SD_scores} and Figure~\ref{fig:contest_res} show that 
for re-identification, 
the standard deviation ($=0.01$) is much smaller than 
the difference between the best privacy score ($=0.79$) and the second-best privacy score ($=0.67$). 
Thus, it is highly unlikely that the difference of the original traces changes the order of the 1st and 2nd places. 
For trace inference, 
the best, second-best, and third-best privacy scores are $0.720$, $0.709$, and $0.637$, respectively. 
Thus, \colorB{the difference of the original traces may affect the order of the 1st and 2nd places. 
However,} it is highly unlikely that the difference of the original traces changes the order of the 2nd and 3rd places. 

\colorB{To improve the fairness (i.e., to further reduce the SD), we should increase the number $m$ of virtual users. 
However, the increase of $m$ would increase the burden of each team during the contest. 
We should consider both the fairness and the burden of each team when we decide a value of $m$.} 

\section{Examples of Sample Attack Algorithms}
\label{app:examples_sample}

\begin{figure}[t]
\centering
\includegraphics[width=0.99\linewidth]{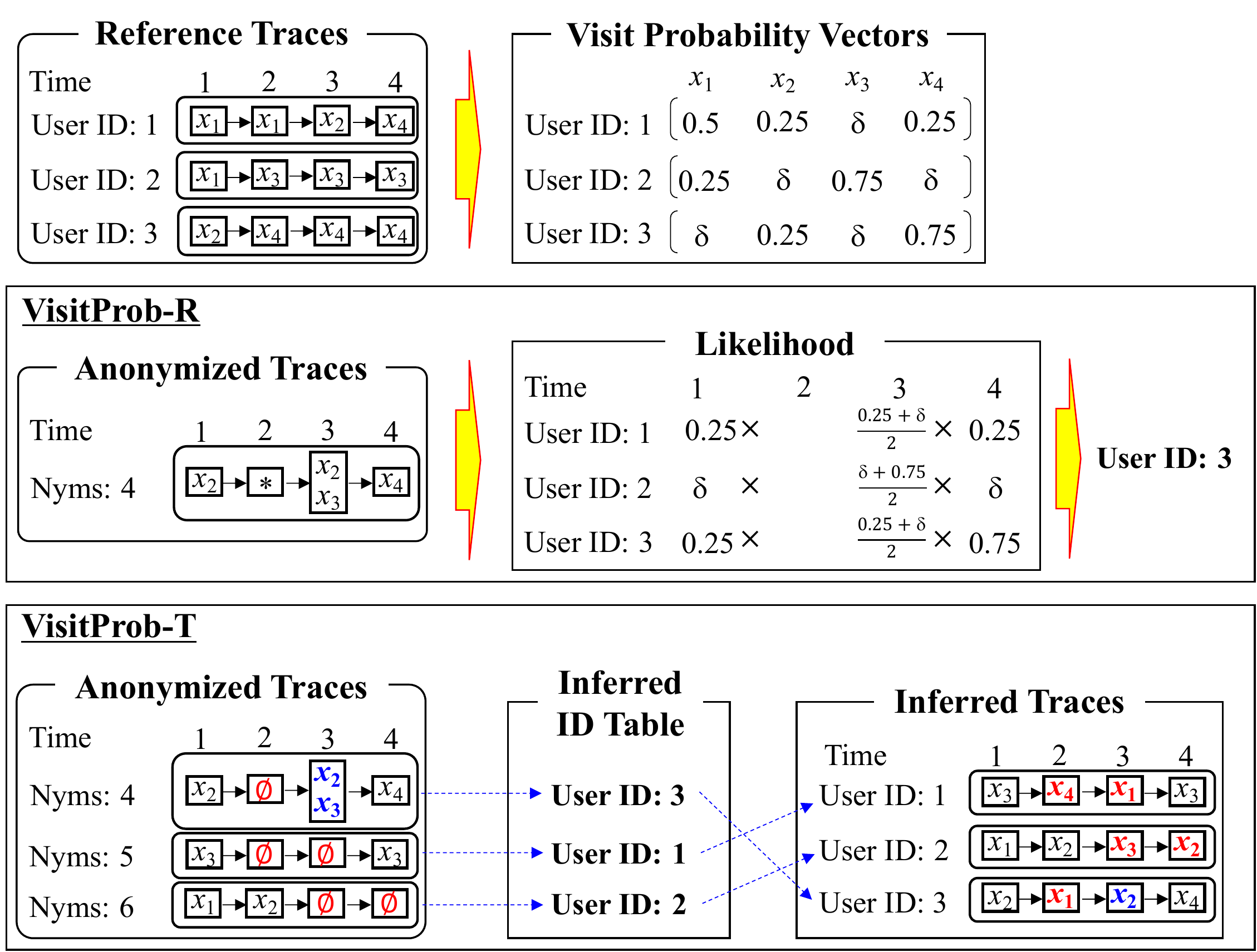}
\conference{\vspace{-3mm}}\arxiv{\vspace{-3mm}}
\caption{\colorB{Examples of visit probability vectors, \textano{VisitProb-R}, and \textano{VisitProb-T}.}} 
\label{fig:smpl_attack}
\end{figure}

The upper panel of Figure~\ref{fig:smpl_attack} shows examples of visit probability vectors. 
Here, we assign a very small positive value $\delta$ ($=10^{-8}$) to an element with zero probability. 
The middle panel of Figure~\ref{fig:smpl_attack} shows an example of \textano{VisitProb-R}. 
In this example, ``user ID = 3'' is output as a re-identification result, as it achieves the highest likelihood. 
The lower panel of Figure~\ref{fig:smpl_attack} shows an example of \textano{VisitProb-T}. 
For perturbation, it outputs the noisy location as is. 
For generalization (marked with blue), it randomly chooses a region from generalized regions. 
For deletion (marked with red), it randomly chooses a region from all regions. 

\end{document}